\documentclass[dvips]{arxstspdf}
\usepackage{graphicx}
\usepackage{flushend,stfloats}

\volume{25}
\issue{3}
\pubyear{2010}
\firstpage{368}
\lastpage{387}
\doi{10.1214/10-STS340}

\makeatletter

\newproclaim{defi}{Definition}
\newtheorem{theo}[defi]{Theorem}
\newtheorem{corol}[defi]{Corollary}

\newproclaim{Example}{Example}
\newproclaim{Example continued}{Example continued}
\newproclaim{remarks}{Remarks}
\setattribute{ead}{sep}{; }
\makeatother

\begin{document}
\begin{frontmatter}

\title{Graphical Models for Inference Under
Outcome-Dependent Sampling}
\runtitle{Graphical Models for Outcome Dependent Sampling}

\begin{aug}
\author[a]{\fnms{Vanessa}
\snm{Didelez}\corref{}\ead[label=e3]{vanessa.didelez@bristol.ac.uk}},
\author[b]{\fnms{Svend} \snm{Kreiner}\ead[label=e1]{s.kreiner@biostat.ku.dk}}
\and
\author[b]{\fnms{Niels} \snm{Keiding}\ead[label=e2]{n.keiding@biostat.ku.dk}}

\runauthor{V. Didelez, S. Kreiner and N. Keiding}

\affiliation{University of Bristol, University of Copenhagen}

\address[a]{Vanessa Didelez is Doctor, Department of Mathematics,
University of Bristol, University Walk, Bristol, BS8 1TW, UK (\printead{e3}).}
\address[b]{Svend Kreiner and Niels Keiding are Professors, Department of
Biostatistics, University of Copenhagen, P.O.B. 2099, 1014 Copenhagen,
Denmark (\printead{e1,e2}).}
\end{aug}

\begin{abstract}
We consider situations where data have been collected such that the sampling
depends on the outcome of interest and possibly further covariates,
as for instance in case-control studies. Graphical models represent
assumptions about the conditional independencies among the variables.
By including
a node for the sampling indicator, assumptions about sampling processes can
be made explicit. We demonstrate how to read off such graphs whether consistent
estimation of the association between exposure and outcome is possible.
Moreover, we give sufficient graphical conditions for testing and
estimating the {\em causal} effect of exposure on outcome.
The practical use is illustrated with a number of examples.
\end{abstract}


\end{frontmatter}

\section{Introduction}

Nonrandom sampling poses a challenge for the statistical
analysis especially of observational data. We focus here on the problem of
outcome-dependent sampling, where the inclusion of a unit into the sample
depends, possibly in some indirect way, on the outcome of interest, and possibly
on further variables. The prime examples are case-control studies,
which have
been surrounded by a long controversy, but are now one of the most popular
designs in observational epidemiology (Breslow, \citeyear{1996Breslow}).
Any observational study based on volunteers is also potentially sampled
depending
on the outcome, as the willingness to
participate can never be safely assumed to be independent of the
outcome of
interest, for example, income.
In other situations the outcome-dependent sampling may not be obvious,
such as
retrospective time-to-event studies (Weinberg, Baird and Rowland, \citeyear{1993Weinberg}).

A superficial statistical analysis will typically be biased under nonrandom
sampling.
It is therefore important to investigate and understand the assumptions and
limitations underlying valid inference in such situations.
Most approaches make very specific parametric
modeling assumptions, including assumptions about the selection mechanism,
sometimes accompanied by a sensitivity analysis (see, e.g., Copas
and Li, \citeyear{1997Copas}, or McCullagh, \citeyear{2008McCullagh}).
As an alternative, in this article we investigate the potential of
graphical models
to address the problem of outcome-dependent sampling and restrict any
assumptions
to be nonparametric and only in terms of conditional independencies. A~graphical model
represents variables as nodes and uses edges between nodes so that
separations reflect conditional independencies in the underlying
model (see, e.g., Whittaker, \citeyear{1990Whittaker}, or Lauritzen, \citeyear{1996Lauritzen}).

A key element of our proposed approach is to include
a separate node as a binary selection indicator in the graph so as to
represent structural assumptions about how the sampling mechanism is related
to exposure, outcome, covariates and possibly hidden variables.
A~similar idea appears in the works of Cooper (\citeyear{1995Cooper}), Cox and Wermuth (\citeyear{1996Cox}),
Geneletti, Richardson and Best  (\citeyear{2009Geneletti}) and Lauritzen and\break Richardson (\citeyear{2008Lauritzen}).
Our main results in this article address the (graphical)
characterization of
situations where the typical null hypothesis of no association or no causal
effect can be tested,
and, when all variables are categorical, where (causal) odds ratios can
be consistently
estimated.
These graphical rules do not require particular parametric constraints
and essentially capture when the model is collapsible over the selection
indicator.
While our results on testing are general,
estimation is restricted to odds ratios as these (or functions thereof)
are the only measures of association that do not depend on the marginal
distributions (Edwards, \citeyear{1963Edwards}; Altham, \citeyear{1970Altham}) and for which results can
be obtained
without specific parametric assumptions.

The outline of the article is as follows.
In Section~\ref{sec_gramo} we review basic concepts of graphical
models, and
highlight how a binary sampling node can be included so as to make assumptions
about the sampling process explicit (Section \ref{subsec_graphs_ods}).
The corresponding graphs can be constructed
in different ways, for instance in a prospective or retrospective manner
representing different types of assumptions.
Section \ref{sec_coll} \mbox{revisits} the notion of collapsibility, which is
fundamental
for being able to `ignore' outcome-dependent sampling. Sections \ref
{subsec_gramo_s} and \ref{sec_test_null} provide the central results
that allow us to estimate an odds ratio (Corollary \ref{corol_retro})
or test for
an association (Theorem \ref{null_test})
under outcome-dependent sampling.
Section \ref{subsec_appl} illustrates this with a data example.
We then move on to causal inference in Section \ref{sec_causal}, where we
define a causal effect as the effect of an intervention. We formalize this
using intervention indicators, and adapt the graphical representation
by adding
a corresponding decision node yielding so-called influence diagrams (Dawid,
\citeyear{2002Dawid}). In analogy to the associational case, Theorem
\ref{theo_prosp} establishes (graphically verifiable) conditions under
which a
{\em prospective} causal
effect can be tested or estimated from {\em retrospective}, that is,
outcome-dependent, data.
In Section \ref{sec_ext} we present new results that apply to less
obvious cases
of outcome-dependent sampling (Theorems \ref{theo1} and \ref{theo5}).

\section{Graphical models}\label{sec_gramo}
We start with a brief overview of graphical models.
Our notation follows closely that of Lauritzen (\citeyear{1996Lauritzen}).
A graph $G=(V,E)$ with nodes or vertices $V$
and edges $E$ is combined with a statistical model, that is, a distribution
$P$ on a set of variables that are identified with the nodes of the graph.
The distribution $P$ has to be such that the {\em absence} of an edge
in the graph
represents a certain {\em conditional independence} between the corresponding
pair of variables. Conditional independence of $A$ and $B$ given $C$
is denoted by $A\perp\!\!\!\perp B|C$ (Dawid, \citeyear{1979Dawid}).

The type of graph dictates the specific conditional independence
induced by the
absence of an edge. A basic distinction is between undirected graphs,
directed acyclic graphs (DAGs) and chain graphs. We review the former
two but do
not go into detail for chain graphs.

\subsection{Undirected Graphs}

\begin{figure}

\includegraphics{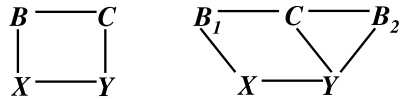}

\caption{Examples of undirected graphs.} \label{fig8}
\end{figure}

In an undirected graph $G$ all edges are undirected, represented by
$a$--$b$. All nodes in $V\backslash\{a\}$ that have an edge with
$a\in V$ form the {\em boundary} $\operatorname{bd}(a)$ of $a$.
We say that two sets of nodes $A$ and $B$ are
{\em separated} by a third set $C$ if any path along the edges of
the graph between $A$ and $B$ includes vertices in $C$.
In particular, each set $A$ is separated by its boundary
from all other nodes $V\backslash(\mathrm{bd}(A)\cup A)$.
The induced conditional independencies are as follows.
For any disjoint subsets $A,B,C\subset V$,
the variables in $A$ have to be conditionally independent of those in
$B$ given $C$ whenever $C$ separates $A$ and $B$ in the graph---we then
call $P$ {\em$G$-Markovian}.
As examples consider the undirected graphs in Figure \ref{fig8}. In
the left graph,
$X\perp\!\!\!\perp C|(B,Y)$ as well as $Y\perp\!\!\!\perp B|(C,X)$.
In the right graph of Figure~\ref{fig8}, for instance,
$B_1\perp\!\!\!\perp B_2|(C,Y)$ or $B_1\perp\!\!\!\perp B_2|(C,X)$
showing that separating sets
are not
necessarily unique.

We say that a subset $C\subset V$ of
the nodes of a graph is {\em complete}
if each pair of nodes in $C$ is joined by an edge. We further call
such a complete $C$ a {\em clique} if adding any further node would
destroy its completeness; that is, a clique is a maximal complete set
of nodes. In Figure \ref{fig8} the graph on the right has cliques
$\{B_1,C\}, \{B_1,X\}, \{X,Y\}$ and $\{B_2,C,Y\}$.

Let $\mathcal{C}$ be the set of all cliques in $G$.
The above conditional independence restrictions that are induced by an
undirected graph
go hand in hand with a factorization of the joint distribution in terms
of these cliques.
Assume $p$ is the p.d.f. (or p.m.f. if discrete) of the random vector
$X_V=(X_1,\ldots, X_K)$ taking values $x_V=(x_1, \ldots,x_K)$;
then $p$ is said to factorize according to an undirected
graph $G$ if it can be written as
%
\begin{equation}\label{gen_fact}
p(x_V)=\prod_{C \in\mathcal{C}} \phi_C(x_C),
\end{equation}
where $\phi_C(\cdot)$ are functions that depend on $x_V$ only through
its components
in $C$.

In later sections we will also make use of the notion of
an induced {\em subgraph} $G_A$,
$A\subset V$, obtained by removing all nodes in $V\backslash A$ and
edges involving at least one node in $V\backslash A$.

\subsection{Directed Acyclic Graphs}

Directed acyclic graphs (DAGs) represent different sets of
conditional independencies than undirected graphs. They often
seem more natural if one thinks of a data generating process, that is,
a way in which the data could be simulated, but DAGs can also be used to
represent conditional independencies other than for a generating
process. Moreover, one could also use chain graphs (Frydenberg, \citeyear{1990aFrydenberg};
Wermuth and Lauritzen, \citeyear{1990Wermuth}) which represent again different sets of conditional
independencies; undirected graphs as well as DAGs are special cases of
chain graphs.

\begin{figure}[b]

\includegraphics{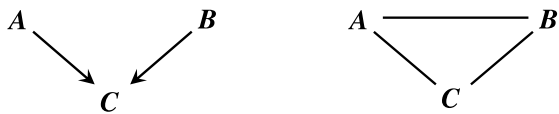}

\caption{DAG (left) representing $A\perp\!\!\!\perp B$ and moral
graph (right)
showing $A\perp\!\!\!\perp\hspace*{-11pt}\diagup\ B|C$.}
\label{selection_DAG}
\end{figure}

In DAGs all edges are directed, for example, $a\rightarrow b$,
without forming
any directed cycles. For $a\rightarrow b$ we say that $a$ is a {\em
parent} of
$b$ and $b$ is a {\em child} of $a$. This can be generalized to sets,
for example, pa($A$), $A\subset V$, denotes all the variables in
$V\backslash A$ that are
parents of some variable in $A$, and similarly for
the children ch$(A)$. Analogously we speak of {\em descendants} de$(A)$ of
$A$, meaning all those nodes in $V\backslash A$ that can be reached
from some vertex in $A$ following
the direction of the edges, while {\em nondescendants} nd$(A)$ are all
other nodes (excluding $A$ itself). Further, the {\em ancestors} an$(A)$
are defined as those nodes in $V\backslash A$ from which we can reach
some vertex in
$A$ following the direction of the edges.

Similarly to (\ref{gen_fact}), a DAG also induces a factorization of the
joint density as follows:
%
\begin{equation}
p(x_1,\ldots,x_K)=\prod_{k=1}^K p\bigl(x_i|x_{\mathrm{pa}(i)}\bigr),
\label{factor}
\end{equation}
where $p(x_i|x_{\mathrm{pa}(i)})$ denotes the conditional density of
$X_i$ given all its parent variables $X_{\mathrm{pa}(i)}$.
A simple example is given in Figure \ref{selection_DAG} (left): here
the joint distribution
factorizes as $p(a,b,c)=p(a)p(b)p(c|a,b)$.

The factorization (\ref{factor}) is equivalent to the following, graphically
characterized conditional independencies:
%
\begin{equation}
X_i\perp\!\!\!\perp X_{\mathrm{nd}(i)\backslash\mathrm
{pa}(i)}|X_{\mathrm{pa}(i)} \quad\forall
i\in V. \label{local_d}
\end{equation}
For instance, in Figure \ref{selection_DAG} (left), $A$ is a nondescendant
of $B$ that has no parents, hence $A\perp\!\!\!\perp B$ is implied by
this DAG;
there is no
other (conditional) independence in this particular DAG.

Even though the partial ordering imposed by the direction of the edges on
the variables $X_1,\ldots,X_K$ is often postulated
to follow some causal or time order, this does not automatically follow
from the
represented conditional independencies (cf.\ Section
\ref{subsec_graphs_ods}).
For instance $p(a,b,c)=p(a)p(c|a)p(b|c)$ implies the same conditional
independencies as $p(a,b,c)=p(b)p(c|b)p(a|c)$, represented in the two
graphs $A\rightarrow C \rightarrow B$ and $A\leftarrow C \leftarrow B$,
respectively.
In both cases $A\perp\!\!\!\perp B|C$.
These two graphs (or factorizations) are called {\em Markov
equivalent}, meaning that exactly the same conditional independencies
can be read off.
This implies that even if we believe there is an underlying unknown
causal or other kind of ordering, then
conditional independencies estimated from observational data on $A,B,C$
cannot help to distinguish between these graphs nor tell us the causal order.
However, depending on the way in which the variables are observed, how
a study is conducted or other considerations, it might seem more
natural to specify $p(c|a)$ and $p(b|c)$ than $p(c|b)$ and $p(a|c)$
(we will come back to this below).
Note that the DAG of Figure \ref{selection_DAG} is not equivalent to
the former two as it induces a different
independence, $A\perp\!\!\!\perp B$.

\subsection{Selection Effect and Moralization}\label{sec_selection}

All conditional independencies that can be deduced from (\ref
{local_d}) are given
by {\em graph separation} for DAGs.
One can either use the {\em d-separation} criterion (Verma and Pearl,
\citeyear{1988Verma}) or
the {\em moralization} criterion (Lauritzen et al., \citeyear{1990Lauritzen}).
The latter is described in detail
next.\footnote{The two criteria, $d$-separation and moralization, are entirely
equivalent and readers more familiar with the former can verify all conditional
independencies in the following with $d$-separation.}

The moralization criterion is used to determine the conditional
independencies of a DAG $G$ in a collection of undirected graphs
using ordinary graph separation. These are the moral graphs on
subgraphs of~$G$. Let $A\subset V$ and let $\operatorname{An}(A)=$ an$(A)\cup A$;
then the corresponding
{\em moral graph} $G_{\mathrm{An}(A)}^m$ is given by adding undirected edges
between nodes of An$(A)$ that have a common child in An$(A)$ and
then turning all remaining directed edges into undirected ones. Any
conditional independence that is induced by factorization (\ref{factor})
can be read
off $G_{\mathrm{An}(A)}^m$ for some $A\subset V$. More specifically,
if we
want to establish whether $A\perp\!\!\!\perp B|C$, then we check for graph
separation in the undirected graph $G_{\mathrm{An}(A\cup B\cup C)}^m$.
In Figure
\ref{selection_DAG}, if we want to\vspace*{1pt} investigate whether $A\perp\!\!\!
\perp B$, we
draw the
moral graph $G_{\mathrm{An}(A\cup B)}^m$ which consists of the two
unconnected nodes $A$
and $B$, with node $C$ removed as it is not in An$(A\cup B)$,
confirming that $A\perp\!\!\!\perp B$ as shown earlier. However, if we
want to check
whether $A\perp\!\!\!\perp B|C$, then we draw the moral graph
$G_{\mathrm{An}(A\cup B\cup
C)}^m$ shown
on the right in\vspace*{1pt} Figure~\ref{selection_DAG} and see that this
independence is not
implied by the graphical model.

The ``moral edges'' represent what is known in epidemiology as
{\em selection} or {\em stratification} (Greenland, \citeyear{2003Greenland};
Hernan, Hern\'andez-D\'iaz and Robins, \citeyear{2004Hern}):
if $A$ and $B$ are marginally independent but $C$ depends on both of
them, as
represented by the DAG in Figure \ref{selection_DAG}, then
conditioning on $C$ typically induces a dependence between $A$ and $B$.
This can easily be seen as the factorization $p(a,b,c)=p(a)p(b)p(c|a,b)$
does {\em not} imply a factorization of $p(a,b|c)=p(a|c)p(b|c)$.
As an example assume that $A$ is (binary) exposure to a risk factor and
$B$ is some disease indicator that is entirely unrelated to $A$.
Further assume that the data are obtained from a database $C$, with
$C=1$ if an individual is found in that database and $C=0$ otherwise.
If, for some reason, individuals who are exposed are more likely to be
in the database as well as individuals who are ill, then
we typically find an association between $A$ and $B$ in the sample from
that database because we condition on $C=1$.
This may, for example, happen when it is a database for a different
disease for which $A$ {\em is} a risk factor and which is associated
with $B$ (cf. Berkson, \citeyear{1946Berkson}).
In this example, the marginal association of $A$ and $B$ is the target
of inference, but cannot be obtained from the available data, so that
this phenomenon is often called selection (or stratification) {\em
bias}. Note that in econometrics the term selection bias is also used
to denote
systematic (as opposed to randomized) selection into treatment or
exposure (Heckman, \citeyear{1979Heckman}), which in epidemiology
would rather be called confounding.

\begin{figure}

\includegraphics{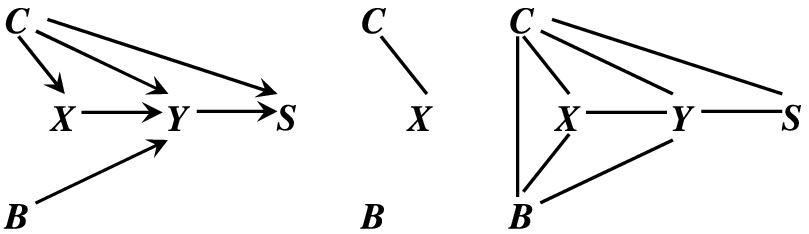}

\caption{DAG \textup{(left)} with moral graphs
on ${\operatorname{An}(B\cup X)}$ \textup{(middle)} and on ${\operatorname
{An}(B\cup X\cup S)}$ \textup{(right)}.}
\label{fig_ccDAG}
\end{figure}

The selection
effect is equally relevant when conditioning is not on a common child
of nodes $A$ and
$B$ but on any descendant of such a common child as this indirectly provides
information on all ancestors; for \mbox{example}, in Figure \ref{fig_ccDAG}
(left), $X\perp\!\!\!\perp B$ but $X\perp\!\!\!\perp\hspace
*{-13pt}\diagup\ B|S$ because $S$ carries some information
on $Y$ and $Y$ is a common child of $B$ and $X$. Figure \ref{fig_ccDAG}
shows the corresponding moral graphs $G_{\mathrm{An}(B\cup X)}^m$
(middle) and
$G_{\mathrm{An}(B\cup X\cup S)}^m$ (right) for checking these two
conditional independence statements.

\subsection{Graphical Representation of Sampling Mechanisms}
\label{subsec_graphs_ods}

We now turn to the question of how to represent with graphical models
that the units in the dataset have possibly been sampled
depending on some of the variables relevant to the analysis.
The nodes include $X$, the exposure or treatment, and $Y$, the response.
Additional nodes are used to represent further relevant variables, for
example, in
particular a binary variable $S$ indicating whether the unit is
sampled, $S=1$,
or not, $S=0$. This use of a sampling indicator
has also been proposed by Cooper (\citeyear{1995Cooper}), Cox and Wermuth (\citeyear{1996Cox}),
Geneletti, Richardson and Best  (\citeyear{2009Geneletti}) and Lauritzen and Richardson (\citeyear{2008Lauritzen}).
Also, we may include a set of covariates $C$.

\begin{figure}
\begin{tabular}{cccc}

\includegraphics{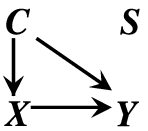}
&\includegraphics{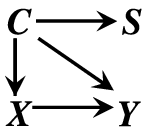}&\includegraphics{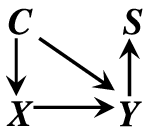}&\includegraphics{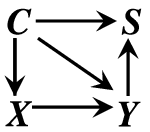}\\
\scriptsize{(a)}&\scriptsize{(b)}&\scriptsize{(c)}&\scriptsize{(d)}
\end{tabular}
\caption{DAGs for different sampling processes: \textup{(a)}
independent sampling,
\textup{(b)} stratified (on covariates $C$) sampling, \textup{(c)}~case-control sampling,
\textup{(d)} case-control matched by $C$.}
\label{sampling_all}
\end{figure}

The graph is normally constructed based on a combination of subject matter
background knowledge, especially concerning the sampling mechanism, and
testable implications.
Some examples for different sampling mechanisms are depicted in Figure
\ref{sampling_all}.
Note that all the graphs in Figure \ref{sampling_all} could as well be
undirected (replacing every directed edge by an undirected one) and
still represent the same conditional independencies, that is, $S\perp\!
\!\!\perp
(X,Y,C)$ in (a), $S\perp\!\!\!\perp(X,Y)|C$ in (b), $S\perp\!\!\!
\perp(X,C)|Y$ in (c) and
$S\perp\!\!\!\perp X|(Y,C)$ in (d).

For a given set of variables we speak of {\em outcome-dependent}
sampling if $Y$ and $S$ are dependent whatever subset of the
remaining variables we condition on, such as in (c) and (d) of Figure
\ref{sampling_all}.
The problem created by outcome-dependent sampling is that, first,
$p(y)$ cannot
be identified as the observations are only informative for $p(y|S=1)$,
and second,
that conditioning on $S=1$ might create associations that are not
present in
the target population due to the selection effect as explained in Section
\ref{sec_selection} (Hernan, Hern\'andez-D\'iaz and Robins, \citeyear{2004Hern}).
For a DAG to represent the selection effect, it has to be
constructed in a {\em prospective} way, as in
Figure~\ref{sampling_all} and as illustrated in the following example.

\begin{Example*} In Figure \ref{fig_ccDAG}, assume that
$B$ and $C$ are baseline covariates like $B={}$sex and $C={}$age, while $X$
is exposure
to a risk factor (e.g., loud music) that possibly changes with age but
not with sex
and $Y$ is a disease (e.g., hearing loss) that is affected by all
previous variables.
Assume further that we select individuals into our study, $S=1$, based
on age and disease
status (as would be the case in a case-control study matched by age);
hence $S$
depends on $C$ and $Y$. A DAG on the partially ordered variables $(\{
B,C\}, X,Y,S)$
would reflect the time order in which the variables are assumed to be realized,
and would enable us to express, for instance, the assumption that
$B\perp\!\!\!\perp X$ as shown
in Figure \ref{fig_ccDAG}.
It also allows us to represent that the sampling induces dependencies
that are not present marginally, that is, the selection effect.
The DAG in Figure \ref{fig_ccDAG}, for instance, implies that the
joint density of all variables factorizes as
\begin{eqnarray*}
&&p(s,y,b,c,x)
\\
&&\quad=p(s|y,c)p(y|x,b,c)p(x|c)p(b)p(c).
\end{eqnarray*}
Data from a case-control study, however, only admits inference on the
conditional
distribution given $S=1$ which is given by
\begin{eqnarray}\label{eq9}
&&p(y,b,c,x|S=1)\nonumber
\\[-8pt]\\[-8pt]
&&\quad=\frac{p(S=1|y,c)p(y|x,b,c)p(x|c)p(b)}{
\sum_{y,c}p(S=1|y,c)p(c) p(y|c)}.\nonumber
\end{eqnarray}
Marginalizing this over $Y$ shows that there are no necessary independencies
among $\{X,B,C\}$ conditional on $S=1$ confirming that $\{X,B,C\}$
must be complete in $G_{\mathrm{An}\{X,B,C,S\}}^m$ as in the right
graph of Figure~\ref{fig_ccDAG}.
\end{Example*}

As an alternative to the {prospective} view, one could decide to represent
the {\em sampling process}, that is, the order imposed by the sampling
which will be {\em retrospective} under outcome-dependent sampling;
in a case-control study, for instance, the response $Y$ is
sampled first and hence the remaining variables are conditional on the
response.

\begin{figure}

\includegraphics{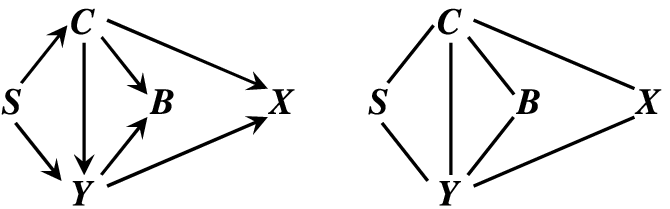}

\caption{Representing the sampling order: DAG \textup{(left)} with
moral graph
on $\operatorname{An}(B\cup X\cup C\cup Y)$ \textup{(right)}.} \label{fig_ccDAG2}
\end{figure}

\begin{Example continued*}
Continuing the above example, choosing the sampled units ($S=1$) based
on age and
disease status partially reverses the order to be $(S, \{C,Y\}, \{B,X\})$.
Conditional independence test on the retrospective data might reveal
that $X\perp\!\!\!\perp B|(C,Y)$ which can be represented as in the DAG
in Figure~\ref{fig_ccDAG2} (cf.\ moral graph on the right). While this
conditional
independence can be tested from case-control data, the
marginal independence $B\perp\!\!\!\perp X$ postulated in Figure~\ref
{fig_ccDAG}
cannot be
tested from case-control data due to the properties of (\ref{eq9}).
(The latter
could, however, be checked approximately, when the disease is rare,
using only the controls.)
\end{Example continued*}

A DAG reflecting the sampling order allows us to encode conditional
independencies given that a unit is sampled.
While this makes it typically more difficult to include any subject-specific
background knowledge about the data generating process in the
formulation of the
graph, it might result in a model that fits the data better and still provides
consistent, and simpler, estimates of the parameter we are interested in.
For a causal analysis, however, where we want to be able to express
prior causal assumptions, it is
crucial that the graph reflects the prospective view (see Section \ref
{ID_CDAG}).
In particular we want to encode which variables are potentially
affected by an
external change in the exposure status $X$ by representing these as
descendants of $X$ in the graph.
The causal interpretation of DAGs will be discussed in more detail
in Section~\ref{sec_causal}; until then we focus on graphs
representing conditional independencies.

\section{Collapsibility}\label{sec_coll}

Roughly speaking, collapsibility means that inference can be carried out
on a subset of the variables, that is, after marginalizing over others
(an exact
definition for odds ratios is given below).
Typically collapsibility is exploited to \textit{reduce} dimensionality and
computational effort, as it enables us to
pool subgroups. In our context collapsibility is relevant for the
opposite reason:
all that we have is a subgroup, namely the sampled population; but
we want the associations found in this subgroup to be valid for the
whole target
population.

In this section we focus on the odds ratio as a measure of association.
This is motivated by the fact that the odds ratio does not depend on the
marginal distribution of $Y$,
which is potentially affected by the sampling process, this also being
the main
reason why odds ratios are typically used for case-control data.
We revisit general results on collapsibility of odds ratios,
including their graphical versions, and then modify these to deal with the
particular problem of outcome-dependent sampling.
Note that the results concerning odds ratios given next are closely
linked to
graphical log-linear models (Darroch et al., \citeyear{1980Darroch}; Lauritzen, \citeyear{1996Lauritzen},
Chapter~4).

\subsection{Conditional Odds Ratios}
Define the conditional odds ratio $\mathit{OR}_{YX}(C=c)$ (in short we also
write $\mathit{OR}_{YX}(C)$)
for binary $Y$ and binary $X$ given $C=c$ as
\[
\frac{p(Y=1|X=1,C=c)p(Y=0|X=0,C=c)}{p(Y=0|X=1,C=c)p(Y=1|X=0,C=c)}.
\]
It is straightforward to generalize this for the case where $Y$ and $X$
have more than
two categories. We might then consider a collection of (conditional)
odds ratios comparing the
probabilities for $Y=y$ versus a reference category $Y=0$ for values
of $X$, say $X=x$ versus $X=0$. This collection of odds ratios fully
characterizes
the (conditional) dependence between $X$ and $Y$ and is, vice versa, fully
determined by the corresponding interaction terms of a log-linear model.
The results given below can therefore easily be extended to the
case of more than two categories.

\subsection{General Results}
It is well known that the conditional odds ratio $\mathit{OR}_{YX}(C)$ is not
necessarily the same as when we collapse over $C$, that is, as the
marginal $\mathit{OR}_{YX}$, even if $\mathit{OR}_{YX}(C=c)=\mathit{OR}_{YX}(C=c')$ for all $c\not=c'$.
Though this property is at the heart of most definitions of
collapsibility, there are some subtleties giving rise to various
definitions of, and different conditions that are sufficient and
sometimes necessary for, collapsibility
(Bishop, Fienberg and Holland, \citeyear{1975Bishop}; Whittemore, \citeyear{1978Whittemore};
Shapiro, \citeyear{1982Shapiro}; Davis, \citeyear{1986Davis};
Ducharme and Lepage, \citeyear{1986Ducharme}; Wermuth, \citeyear{1987Wermuth};
Geng, \citeyear{1992Geng}; Guo, Geng and Fung, \citeyear{2001Guo}).
Here we define collapsibility as follows.

\begin{defi}
Consider two binary variables $X$ and $Y$ and disjoint sets of further variables
$B$ and $C$. We say that the odds ratio $\mathit{OR}_{YX}(B,C)$ given $B$ and
$C$ is
collapsible over $B$ if $ \mathit{OR}_{YX}(B=b,C=c)=\mathit{OR}_{YX}(B=b',C=c)=$
$\mathit{OR}_{YX}(C=c)$, for all $ b\not=b'$.
\end{defi}

Note that in the above definition as well as in all the following
results, the covariates $C$ can be of arbitrary measurement level,
while $X$, $Y$ and the covariates we consider to collapse over, $B$,
are categorical.
Collapsibility can then be ensured as follows.

\begin{theo}\label{theo2}
Sufficient conditions for the conditional odds ratio $\mathit{OR}_{YX}(B,C)$ to
be collapsible over $B$ are:
\begin{longlist}[(ii)]
\item[(i)] $Y\perp\!\!\!\perp B|(C,X)$ or
\item[(ii)] $X\perp\!\!\!\perp B|(C,Y)$.
\end{longlist}
\end{theo}

\begin{pf}
This follows from the work of Whittemore (\citeyear{1978Whittemore}).
\end{pf}

\begin{remarks*}
(a) The conditions in Theorem \ref{theo2} are necessary if $B$ is a
single \textit{binary} variable (Whittemore, \citeyear{1978Whittemore}).

(b) The conditions in Theorem \ref{theo2} also ensure, and are necessary
for, \textit{strong} collapsibility which posits that the equality holds
for any newly
defined $B'$ obtained by merging categories of $B$ (Ducharme and
Lepage, \citeyear{1986Ducharme}; Davis, \citeyear{1986Davis}).
\end{remarks*}

The conditions in Theorem \ref{theo2} are not necessary; the following
corollary
gives a more general result.

\begin{corol}\label{corol_succ}
Assume that $B$ can be partitioned into $(B_1, \ldots, B_K)$, and let
$\bar B^{k+1}=(B_{k+1},\ldots,B_K)$. If $B_k$
satisfies for each $k=1,\ldots,K$, either:
\begin{longlist}[(ii)]
\item[(i)] $Y\perp\!\!\!\perp B_k|(C,X,\bar B^{k+1})$ or
\item[(ii)] $X\perp\!\!\!\perp B_k|(C,Y,\bar B^{k+1})$,
\end{longlist}
then $\mathit{OR}_{YX}(B,C)$ is collapsible over $B$.
\end{corol}

\begin{pf}
With Theorem \ref{theo2}, $\mathit{OR}_{YX}(B_k,\ldots,B_K,C)$ is collapsible over
$B_k$, $k=1,\ldots,K$. Hence we can consecutively collapse over $B_1,
\ldots, B_K$.
\end{pf}

The conditions of Theorem \ref{theo2}, generalized in Corollary \ref{corol_succ},
can be checked graphically as they
correspond to simple separations in graphical models, regardless
whether an
undirected graph, a DAG or a chain graph is used to model the data.
We consider the case of undirected graphs next; these could also be
the moral graphs derived from DAGs or chain graphs.

\begin{figure}[b]

\includegraphics{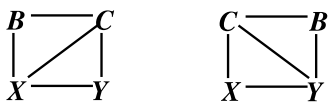}

\caption{Left graph satisfies \textup{(i)}, right graph satisfies
\textup{(ii)} of
Corollary \protect\ref{corol_succ}.} \label{corol3_ill}
\end{figure}

\begin{figure}[t]
\begin{tabular}{cccc}

\includegraphics{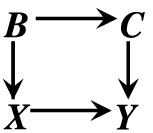}
&\includegraphics{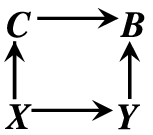}&\includegraphics{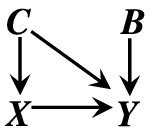}&\includegraphics{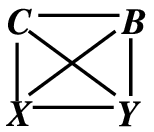}\\
\scriptsize{(a)}&\scriptsize{(b)}&\scriptsize{(c)}&\scriptsize{(d)}
\end{tabular}
\caption{Graphs \textup{(a)} and \textup{(b)} satisfy Theorem \protect\ref{theo2},
while graph \textup{(c)} violates
the conditions [moral graph in \textup{(d)}].} \label{corol3_ill_dag}
\end{figure}

\begin{Example*}
When $B$ is not partitioned, the graphical equivalents of
the two conditions in the above corollary are given in Figure \ref{corol3_ill},
where each of the graphs could have fewer edges but not more. An
example where $B$
consists of $B=\{B_1, B_2\}$ and is collapsible is given in Figure \ref{fig8}
(right); $B_1$ satisfies (i) and $B_2$ satisfies (ii) of Corollary
\ref{corol_succ}.
The conditions of Theorem \ref{theo2} can easily be checked on DAGs as
well. For
instance, in Figure \ref{corol3_ill_dag}, the graph in (a) satisfies
(i) and
(b) satisfies (ii) of the theorem
(their moral graphs are exactly those in Fig. \ref{corol3_ill}). In
contrast, (c) cannot be collapsed over $B$
as the moral graph in (d) shows that neither $B\perp\!\!\!\perp
Y|(X,C)$ nor $B\perp\!\!\!\perp
X|(Y,C)$ holds in general.
Note that the marginal independence $X\perp\!\!\!\perp B$ in this DAG
does not help
with respect to collapsing the odds ratio over $B$.
\end{Example*}

Theorem \ref{theo2} implies that even if we ignore $B$ we can still
obtain \textit{consistent} estimates for
the conditional odds ratio. However, it does not ensure that the actual
value of the
ML-estimate of $\mathit{OR}_{YX}(C)$ is the same in the model where $B$ is
ignored as opposed to when $B$ is included; this is another type of
collapsibility (cf.  Asmussen and Edwards, \citeyear{1983Asmussen}, Lauritzen, \citeyear{1982Lauritzen},
and the discussion by Kreiner, \citeyear{1987Kreiner};
for DAGs see Kim and Kim, \citeyear{2006Kim} and Xie and Geng, \citeyear{2009Xie}; for chain
graphs Didelez and Edwards, \citeyear{2004Didelez}).

\subsection{Collapsibility Under Outcome-Dependent Sampling}\label{subsec_gramo_s}

Now, we investigate collapsibility over $S$ because
in the available data we have $S=1$, so all we can estimate is
necessarily conditional on $S=1$.
Hence, we want to ensure that our estimate for $\mathit{OR}_{YX}(C,S=1)$
applies to the whole target population, that is, is consistent for $\mathit{OR}_{YX}(C)$.

\begin{corol}\label{corol_retro}
The conditional odds ratio\break $\mathit{OR}_{YX}(C,S)$ is collapsible over $S$
if and only if $Y\perp\!\!\!\perp\break S |(C,X)$ or $X\perp\!\!\!\perp S|(C,Y)$.
\end{corol}

\begin{pf}
This follows from Theorem \ref{theo2} and note (a) (see Whittemore,
\citeyear{1978Whittemore}).
\end{pf}

Similarly to Geneletti, Richardson and Best (\citeyear{2009Geneletti}), we can call a set of variables
$C$ satisfying
Corollary \ref{corol_retro} \textit{bias-breaking} because
it allows us to estimate $\mathit{OR}_{YX}(C)$ consistently.
As addressed in Section \ref{sec_selection}, conditioning on $S$ when
no such $C$
can be found and $S$ depends
on both $X$ and $Y$ will typically induce an association even when
there is no
association between $X$ and $Y$, marginally or conditionally on covariates;
and if there is an association between $X$ and $Y$,
then conditioning on $S$ will typically change this association so that
estimates based on the selected data may be biased.

\begin{Example continued*} In both DAGs, Figures \ref{fig_ccDAG} and
\ref{fig_ccDAG2},
we can collapse $\mathit{OR}_{YX}(B,C,S)$ over $S$ as\break $X\perp\!\!\!\perp S|(B,C,Y)$.
We can also collapse, in both\break graphs, $\mathit{OR}_{YX}(C,S)$ over $S$ as
$X\perp\!\!\!\perp
S|(C,Y)$.
\end{Example continued*}

\begin{figure}[t]

\includegraphics{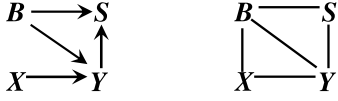}

\caption{DAG for a matched (on $B$) case-control study \textup{(left)} and moral
graph \textup{(right)}.}
\label{matched}
\end{figure}

A consequence of Corollary \ref{corol_retro} is that in a typical
matched case-control
study, the exposure-response odds ratio is only collapsible over the
sampling if
we condition on the matching variables, even if these are not
marginally associated with
exposure. This is illustrated in Figure \ref{matched}, where sampling
depends on the outcome
$Y$ as well as on a matching variable $B$, while $X\perp\!\!\!\perp
B$. Here, we
cannot collapse over $S$ if $B$
is ignored, as neither $X\perp\!\!\!\perp S|Y$ (as can be seen from
the moral
graph, where
the common child $Y$ induces an additional edge between $X$ and $B$
opening a path
to $S$) nor obviously $Y\perp\!\!\!\perp S|X$. However, $X\perp\!\!\!
\perp S|(B,Y)$ so that the
odds ratio
is collapsible over $S$ if it is conditional on $B$.

In addition to collapsing over $S$, we might want to reduce
dimensionality of
covariates, for example, to improve stability of estimates of odds
ratios (Robinson and
Jewell, \citeyear{1991Robinson}).
This is possible if the set of covariates can be written as $(B,C)$
such that $\mathit{OR}_{YX}(B,C,S)$ is collapsible over $S$ and $B$
in a way such that an estimate for $\mathit{OR}_{YX}(C,S=1)$ is consistent for
$\mathit{OR}_{YX}(B,C)$.
The following is straightforward from Theorem \ref{theo2} and assumes
that there is outcome-dependent sampling, that is, $Y\perp\!\!\!\perp
\hspace*{-13pt}\diagup\ S|(B,C,X)$
so that, unlike Corollary \ref{corol_retro}, the next corollary is not
symmetric in $X$ and $Y$.

\begin{corol}\label{cor1}
The odds ratio $\mathit{OR}_{YX}(B,C,S)$ is collapsible over $S$, over $(B,S)$
and over $B$
if $X\perp\!\!\!\perp S|(Y,B,C)$ and:
\begin{longlist}[(ii)]
\item[(i)]
$X\perp\!\!\!\perp B|(Y,C)$ or
\item[(ii)]
$Y\perp\!\!\!\perp B|(X,C)$ and $X\perp\!\!\!\perp S|(Y,C)$.
\end{longlist}
\end{corol}

\begin{pf}
First note that $X\perp\!\!\!\perp S|(Y,B,C)$ yields
$\mathit{OR}_{YX}(B,C,S)$ collapsible over $S$.
For part (i) additionally, $X\perp\!\!\!\perp B|(Y,C)$ yields
$\mathit{OR}_{YX}(B,C)$ collapsible over $B$ by Theorem \ref{theo2}. Both conditional
independencies together imply that $X\perp\!\!\!\perp S|(Y,C)$ which
finally yields
$\mathit{OR}_{YX}(C,S)$ collapsible over $S$, so that all these are equal to
$\mathit{OR}_{YX}(C)$.
For part (ii) we see that $\mathit{OR}_{YX}(B,C)$ is collapsible over $B$ due to
$Y\perp\!\!\!\perp B|(X,C)$ and we further have that $\mathit{OR}_{YX}(C,S)$ is
collapsible
over $S$
due to $X\perp\!\!\!\perp S|(Y,C)$. This yields the desired result.
\end{pf}

\begin{figure}[t]

\includegraphics{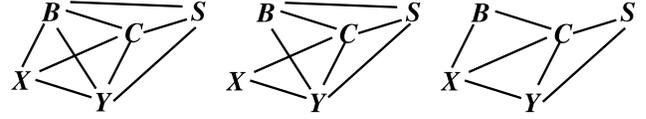}

\caption{Illustrations of Corollary \protect\ref{cor1}: left graph satisfies
$X\perp\!\!\!\perp S|(Y,B,C)$, middle graph satisfies
condition \textup{(i)} and right graph satisfies condition \textup{(ii)}.}
\label{fig_corol1}
\end{figure}

The conditions of Corollary \ref{cor1} can again be checked on graphical
models by corresponding separations; see Figure \ref{fig_corol1}.
As before, if $B$ can be appropriately partitioned,
Corollary \ref{cor1} can be applied successively to the subsets $B_k$.
When DAGs are used to check Corollary \ref{cor1}, then the moral
graph(s) have to be identical to or
have fewer edges than those in Figure \ref{fig_corol1}.
The DAGs in Figures \ref{fig_ccDAG} and \ref{fig_ccDAG2} serve as
examples for
the prospective and retrospective approaches, respectively.
From both we infer that we can collapse over $S$, but only the second
one also
satisfies (i) of Corollary \ref{cor1}.
In contrast it can be seen from Figure \ref{fig_ccDAG}
that the conditions (i) and (ii) of Corollary \ref{cor1}
will not be satisfied in a DAG that represents the prospective view and
where $B$ as well as $X$ point at $Y$.

\subsection{Testability of the Null Hypothesis}\label{sec_test_null}

In general data situations, for example, when $Y$ is continuous, the
odds ratio is not
necessarily an appropriate measure of dependence. Other measures are
typically not
identified under outcome-dependent sampling without further assumptions.
However, a result analogous to Theorem \ref{theo2}
can still be obtained if we restrict ourselves
to the question whether the (conditional) independence of $X$ and $Y$, possibly
given covariates $C$, can still be tested under outcome-dependent sampling.
This conditional independence is often the null hypothesis of interest.

\begin{theo}\label{null_test}
If $S\perp\!\!\!\perp Y|(C,X)$ or $S\perp\!\!\!\perp X|(C,Y)$, then
\[
Y\perp\!\!\!\perp X|C \quad\Longleftrightarrow\quad Y\perp\!\!\!\perp X|(C,S=1).
\]
\end{theo}

\begin{pf}
By the properties of conditional independence (Dawid, \citeyear{1979Dawid}) we have that
$Y\perp\!\!\!\perp X|C$ together with $S\perp\!\!\!\perp Y|(C,X)$ (or
$S\perp\!\!\!\perp X|(C,Y)$) immediately
implies $Y\perp\!\!\!\perp X|(C,S)$. Now assume that\break $Y\perp\!\!\!
\perp X|(C,S=1)$ and $S\perp\!\!\!\perp Y|(C,X)$;
then $p(y|x,c)=\sum_{s}p(y|s,x,c)p(s|x,c)=
\sum_{s}p(y|S=1,c)p(s|x,c)$\break which is just $p(y|S=1,c)$. If instead
$S\perp\!\!\!\perp X|(C,Y)$, an analogous argument yields
$p(x|y,c)=p(x|S=1,c)$.
This completes the proof.
\end{pf}

Hence, under the assumptions of Theorem \ref{null_test} we can test
the null
hypothesis
of no (conditional) association between exposure and response even
under outcome-dependent sampling by any appropriate test for $Y\perp\!
\!\!\perp X|(C,S=1)$.
Note that $S$ has to satisfy the same conditions as for collapsibility
of the
odds ratio in Theorem \ref{theo2}; this can be explained by the one-to-one
relation between (conditional) independence
and vanishing mixed derivative measures of interaction
(Whittaker, \citeyear{1990Whittaker}, page 35) which are a generalization of odds ratios to
continuous distributions.

\subsection{Application: Hormone Replacement Therapy (HRT) and
Transient Cerebral Ischemia (TCI)}\label{subsec_appl}

We illustrate the above with a simplified version of the analysis of
Pedersen et al. (\citeyear{1997Pedersen}). The data are from a case-control study,
where the disease of interest is transient cerebral ischemia ($\mathit{TCI}$)
and the main risk factor is
use of hormone replacement therapy ($\mathit{HRT}$). Controls are matched by age.
Further, smoking status ($\mathit{Smo}$), occupation ($\mathit{Occ}$) and history of other
thromboembolic disorders ($\mathit{THist}$) are included.
All covariates here are categorical; in particular
$\mathit{HRT}$ is measured with categories
``never'' (the reference category), ``former,'' ``oestrogen'' and ``combined.''
The target of inference is the $\mathit{TCI}$--$\mathit{HRT}$ odds ratio conditional on
\textit{all}
covariates. Under what additional assumptions this can be given a
causal interpretation will be addressed explicitly
in the next section.

\begin{figure}[b]

\includegraphics{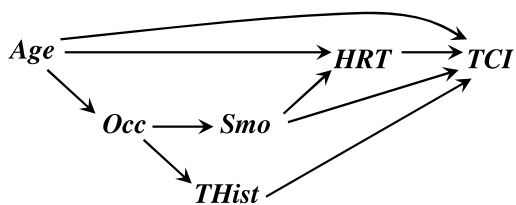}

\caption{Prospective model for $\mathit{HRT}$--$\mathit{TCI}$ example.}
\label{fig_tcig}
\end{figure}

\begin{figure}[t]

\includegraphics{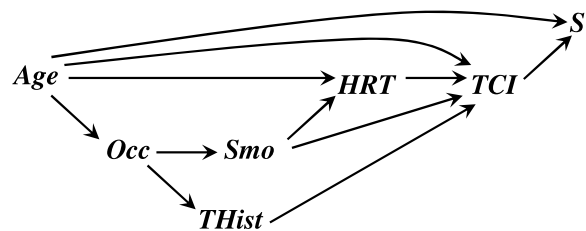}

\caption{DAG as in Figure \protect\ref{fig_tcig} but now including selection
node $S$ to reflect matched case-control sampling.}
\label{fig_tci1}
\end{figure}

Assume the conditional independencies represented in the DAG in Figure
\ref{fig_tcig}.
The additional knowledge that the actual study design is case-control
matched by age is easily included by drawing
arrows from $\mathit{Age}$ and $\mathit{TCI}$ into the additional node $S$, as in Figure
\ref{fig_tci1}.
Note that the assumptions implied by the subgraph on the covariates are
supported by the data from the controls only and extrapolated to hold
for the whole population.

\begin{figure}[b]

\includegraphics{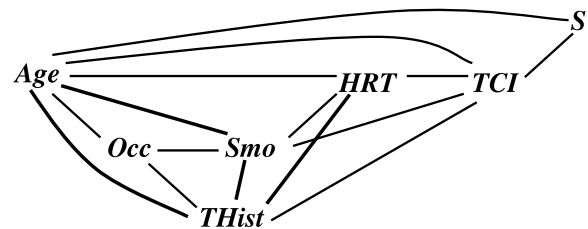}

\caption{Moral graph on all nodes of Figure \protect\ref{fig_tci1}.}
\label{fig_tci1m}
\end{figure}

The moral graph on all variables is shown in Figure \ref{fig_tci1m}.
This represents
the conditional independence structure that we would expect to see in the
data (i.e., conditional on $S=1$).
Note that a particular feature of the age matching is that $\mathit{TCI}\perp\!
\!\!\perp
\mathit{Age}|S=1$ but as this is not necessarily the case for $S=0$ we have to
leave the edge $\mathit{TCI}$--$\mathit{Age}$ in the moral graph.
(It is clear that the $\mathit{TCI}$--$\mathit{Age}$ odds ratio cannot be estimated
from case-control data matched by age; formally, it is not collapsible
over $S$.)

It is obvious from the moral graph as well as from the design of the study
that, given $\mathit{TCI}$ and $\mathit{Age}$, all other variables are independent of the sampling
indicator $S$. In particular $\mathit{HRT} \perp\!\!\!\perp S|(\mathit{TCI},
\mathit{Age}, \mathit{Occ},\break
\mathit{Smo}, \mathit{THist})$
so that Corollary \ref{corol_retro} is satisfied, meaning that the
conditional $\mathit{TCI}$--$\mathit{HRT}$ odds ratio (given all covariates)
based on the selected sample is consistent for the one in the population.
In addition we see that
$B=\mathit{Occ}$ is independent of $\mathit{TCI}$ given $\mathit{Age}$, $\mathit{Smo}$, $\mathit{THist}$ and $\mathit{HRT}$
so that
condition (ii) of Corollary \ref{cor1} holds and we can ignore
the occupation of a person when estimating the conditional odds ratio between
$\mathit{TCI}$ and $\mathit{HRT}$.

The actual calculation of the desired odds ratio can be carried out
by fitting a log-linear model on the subgraph of Figure \ref
{fig_tci1m} excluding the
selection node $S$ (Darroch, Lauritzen and Speed, \citeyear{1980Darroch}; Lauritzen, \citeyear{1996Lauritzen}, Chapter 4).
The desired conditional $\mathit{TCI}$--$\mathit{HRT}$ odds ratio is a function of the
interaction parameters in this model.
For this dataset, we obtain the log (conditional) odds ratios given in
Table \ref{table1}
(there is no evidence that these are different in the subgroups defined
by the conditioning variables
$\mathit{Age}$, $\mathit{THist}$, $\mathit{Smo}$).

\begin{table}[t]
\caption{Conditional odds ratio between $\mathit{HRT}$ and $\mathit{TCI}$
given $\{\mathit{Age}$, $\mathit{THist}$, $\mathit{Smo}\}$
based on separations in the DAG of Figure \protect\ref{fig_tci1}}\label{table1}
\begin{tabular*}{\tablewidth}{@{\extracolsep{4in minus 4in}}lcc@{}}
\hline
$\bolds{\mathit{HRT}}$ \textbf{level} & \textbf{log-\textit{OR} (stdev)} & \textbf{\textit{OR}}\\
\hline
Never & Reference & \\
Former & 0.64 (0.16) & 1.90\\
Oestrogen & 0.73 (0.21) & 2.07\\
Combined & 0.26 (0.19) & 1.29\\
\hline
\end{tabular*}
\end{table}

\begin{figure}[b]

\includegraphics{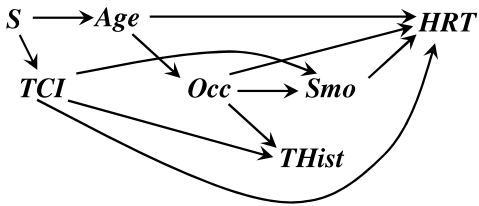}

\caption{DAG reflecting the sampling process for the $\mathit{HRT}$--$\mathit{TCI}$ example.}
\label{fig_tci3}
\end{figure}

Earlier we assumed that the conditional $\mathit{TCI}$--$\mathit{HRT}$
odds ratio given \textit{all other covariates} is the target of interest.
If for some reason instead one wants to condition only on a subset of
$\{\mathit{Occ}, \mathit{Smo}, \mathit{THist},\break \mathit{Age}\}$, this is still collapsible over $S$ as long
as $\mathit{Age}$ is included in that
subset.

An alternative approach is to use a DAG that factorizes retrospectively
according to the sampling process so that $S$
(and hence $\mathit{TCI}$ and $\mathit{Age}$) are the initial variables, taking into
account that observations are conditional on being sampled in the first place.
Assume the conditional independencies represented in the DAG in Figure
\ref{fig_tci3} which is supported by the data.
Collapsibility over $S$ (given the covariates) is of course still
satisfied as this is implied by the
design and still reflected in the model assumptions encoded by the graph.

\begin{figure}[t]

\includegraphics{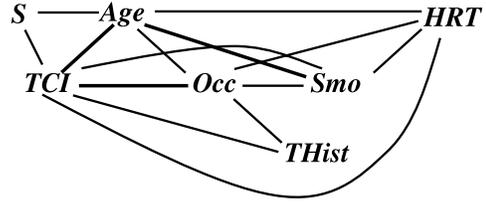}

\caption{Moral graph for Figure \protect\ref{fig_tci3}.}
\label{fig_tci3m}
\end{figure}

From the moral graph in Figure \ref{fig_tci3m} we now have that
$\mathit{HRT}$ is conditionally independent of $\mathit{THist}$\break given the remaining variables
so that condition (i) of Corollary~\ref{cor1} is satisfied with $B=\mathit{THist}$.
The conditional $\mathit{TCI}$--$\mathit{HRT}$ odds ratio given $\mathit{Occ}$, $\mathit{Smo}$, $\mathit{Age}$
estimated from the
case-control data is now consistent for the desired odds ratio in the target
population.
The results are similar to the first model as can be seen from Table \ref{log-or-table2}.
They are not exactly the same as the model assumptions of Figures \ref
{fig_tci1} and
 \ref{fig_tci3} are indeed different, but they are both
consistent, under their respective model, for the
same odds ratio given \textit{all} covariates.

\begin{table}
\caption{Conditional odds ratio between $\mathit{HRT}$ and $\mathit{TCI}$
given $\{\mathit{Age}$, $\mathit{Occ}$, $\mathit{Smo}\}$
based on separations in the DAG of Figure \protect\ref{fig_tci3}}\label{log-or-table2}
\begin{tabular*}{\tablewidth}{@{\extracolsep{4in minus 4in}}lcc@{}}
\hline
$\bolds{\mathit{HRT}}$ \textbf{level} & \textbf{log-\textit{OR} (stdev)} & \textbf{\textit{OR}}\\
 \hline
Never & Reference & \\
Former & 0.66 (0.16) & 1.93\\
Oestrogen & 0.74 (0.21) & 2.10\\
Combined & 0.28 (0.19) & 1.32\\
\hline
\end{tabular*}
\end{table}

\begin{figure}[b]

\includegraphics{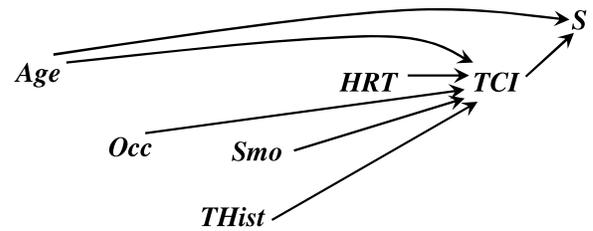}

\caption{Graphical assumptions implicit in a logistic regression.}
\label{fig_logistic}
\end{figure}

\begin{figure}[b]

\includegraphics{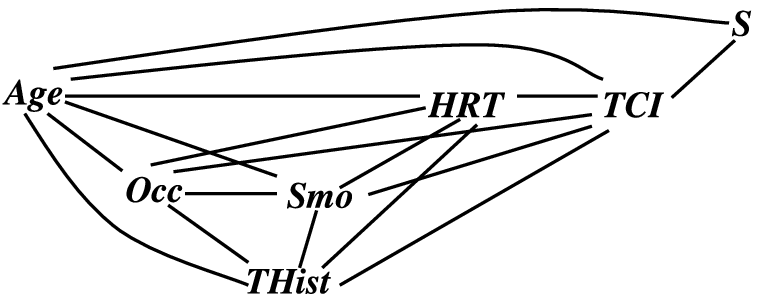}

\caption{Moral graph for Figure \protect\ref{fig_logistic}.}
\label{fig_logistic_m}
\end{figure}

A standard analysis based on a logistic regression of $\mathit{TCI}$ on
explanatory variables
$\mathit{Age}, \mathit{Occ}, \mathit{Smo},\break \mathit{THist}, \mathit{HRT}$ implicitly assumes the model in Figure
\ref{fig_logistic}, that is, all covariates are parents of $\mathit{TCI}$.
While the logistic regression does not make
assumptions about the relations between
the covariates, we have drawn the graph assuming they are mutually
independent. This is to demonstrate that the moral graph, given in Figure
\ref{fig_logistic_m}, in any case has all covariates forming a
complete subgraph, that is, there are no conditional independencies
given $\mathit{TCI}$.
The results can therefore be different from the above analyses, as
conditional independencies involving the covariates cannot be exploited to
collapse over either $\mathit{Occ}$ or $\mathit{THist}$. Adjusting for more covariates
than necessary can lead to larger standard errors in logistic regressions
(see Robinson and Jewell, \citeyear{1991Robinson}), but this happens not to be the case here;
see Table \ref{log-or-table3}.

In a more realistic analysis there will be
more variables to be taken into account, such as menopause, other
medical conditions (hypertension, diabetes,\break heart diseases) and body
mass index
(see Pedersen et al., \citeyear{1997Pedersen}),
so that logistic regression produces larger standard errors. The
graphical approach based on Corollary~\ref{cor1} can help to reduce
the set of
covariates to be adjusted for.

\begin{table}[t]
\caption{Conditional odds ratio between $\mathit{HRT}$ and $\mathit{TCI}$
given $\{\mathit{Age}$, $\mathit{Occ}$, $\mathit{Smo}$, $\mathit{THist}\}$
from a logistic regression}\label{log-or-table3}
\begin{tabular*}{\tablewidth}{@{\extracolsep{4in minus 4in}}lcc@{}}
\hline
$\bolds{\mathit{HRT}}$ \textbf{level} & \textbf{log-\textit{OR} (stdev)} & \textbf{\textit{OR}}\\
\hline
Never & Reference & \\
Former & 0.66 (0.16) & 1.93\\
Oestrogen & 0.76 (0.21) & 2.14\\
Combined & 0.28 (0.19) & 1.32\\
\hline
\end{tabular*}
\end{table}

\section{Causal Effects of Interventions}\label{sec_causal}

So far we have regarded the conditional odds ratio given all covariates
as the target measure of association
between $X$ and $Y$. However, in many situations one is interested
in the  \textit{causal effect} of $X$ on $Y$, not just the association.
A causal effect is
meant to represent the effect that  \textit{manipulations or interventions}
 in $X$
have on~$Y$, as opposed to the mere observation of different $X$ values.
Hence we define the causal effect formally as the effect of an intervention.
Our approach goes back to the work of Spirtes, Glymour and Scheines (\citeyear{1993Spirtes}), Pearl
(\citeyear{1993Pearl}) and is
detailed in the article by Dawid (\citeyear{2002Dawid}) (see also Lauritzen, \citeyear{2000Lauritzen};
Dawid and Didelez, \citeyear{2005Dawid}).
We define an indicator $\sigma_X$ for an intervention in~$X$, where
$\sigma_X$ indicates either that $X$ is being set to a value
$x\in\mathcal{X}$ in the domain of $X$, or that $X$ arises naturally. In
the former case we write $\sigma_X=x$, $x\in\mathcal{X}$, and in the
latter $\sigma_X=\varnothing$. More precisely,
%
\begin{equation}
p(x'|W;\sigma_X=x)=\delta\{x=x'\}, \label{eq10}
\end{equation}
where $W$ can be any set of additional variables and $\delta$ is the
indicator function.
Hence $X$ is independent of any other variable when $\sigma_X=x$. In contrast,
$p(x'|W;\sigma_X=\varnothing)$ is the conditional distribution of
$X$ given $W$ that we observe when no intervention takes place,
that is, if $X$ arises naturally. More generally one may be interested
in other types of interventions, for example, where (\ref{eq10}) is a
probability or depends on $W$ (Dawid and Didelez, \citeyear{2005Dawid}; Didelez et
al., \citeyear{2006Didelez}), but we do not consider these in more detail here.
The above approach is related to the potential outcomes framework
(Rubin, \citeyear{1974Rubin},
\citeyear{1978Rubin}; Robins, \citeyear{1986Robins}), in that the distribution of the outcome $Y$ under an
intervention, $p(y|\sigma_X=x)$, corresponds to the distribution of
the potential outcome~$Y_x$. A comparison of different causal frameworks
can be found in the work of Didelez and Sheehan (\citeyear{2007bDidelez}).
We also call the situation $\sigma_X=\varnothing$ the \textit{observational regime}
and the situation $\sigma_X=x$, for some $x\in\mathcal{X}$, the {\em
experimental} or
\textit{interventional regime}.

\subsection{Influence Diagrams and Causal DAGs}\label{ID_CDAG}

\begin{figure}[b]

\includegraphics{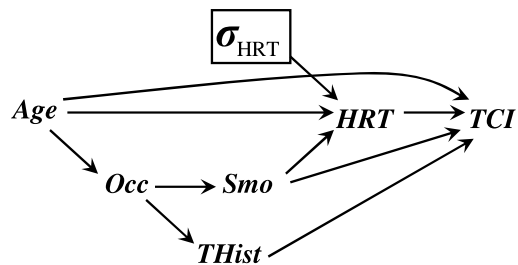}

\caption{Influence diagram for $\mathit{TCI}$ example, prospectively.}
\label{fig_tci2}
\end{figure}

The indicator $\sigma_X$ must be regarded as a decision variable or
parameter, not as a random variable and hence every
statement about the system under investigation must be made
conditional on the value of $\sigma_X$. We will use
conditional independence statements of the type ``$A$ is independent of
$\sigma_X$
given $B$,'' or $A\perp\!\!\!\perp\sigma_X|B$,
meaning that the conditional distribution of $A$ given $B$ is
the same under observation and any setting of $X$.
With this notion of conditional independence applied to the
intervention indicator,
we can then also include $\sigma_X$
into our DAG representation of a data situation in order to encode which
variables are conditionally independent of $\sigma_X$ in the above sense.
As $\sigma_X$ is not a random
variable but a decision variable it is graphically represented in a box and
the resulting DAG is called an influence diagram (Dawid, \citeyear{2002Dawid});
cf. Figure~\ref{fig_tci2} for an example.

The following points are important when constructing an influence
diagram.

(1)
As the decision to intervene in $X$ immediately affects its distribution,
$\sigma_X$
has to be a graph parent of $X$, while $\sigma_X$ itself has no
parents as it is a
decision node.

(2)
Hence, any variables that are nondescendants
of $\sigma_X$ are assumed independent of $\sigma_X$, that is, they are
not affected by an intervention in $X$. Such variables are often called
``pre-treatment'' or ``baseline'' covariates, such as age, gender, etc.
Figure \ref{fig_tci2}, for instance, encodes the assumption that the
distribution
of age, occupation,
smoking status and prior thromboembolic disorders does not change if
HRT is
manipulated, while TCI is potentially affected.
Thus, by representing variables as nondescendants or descendants of
$\sigma_X$
we can explicitly distinguish between variables that
are known a priori not to be affected by an intervention in $X$ and
those that are.
It is therefore not sensible to add such an intervention node
to a retrospective graph such as Figure~\ref{fig_tci3} as important prior
knowledge about what is and is not potentially affected by an
intervention in $X$
could then not be represented.
Retrospective graphs encode a different set of assumptions that
can be used to justify collapsibility as illustrated in Section \ref
{subsec_appl}
in order to apply condition (i) of Corollary \ref{cor1}, for
instance.

(3)
Finally, $X$ being the only child of $\sigma_X$ encodes the assumption that
variables that are potentially affected by an intervention (i.e.,\ descendants
of $X$) are conditionally independent of $\sigma_X$ given ($X$, pa$(X)$).
Justification of this assumption requires us to makes the system ``rich'' enough,
often by including unobservable variables.
Figure \ref{fig_tci2} assumes that $\mathit{TCI}\perp\!\!\!\perp
\sigma_{\mathit{HRT}}|(\mathit{Age}, \mathit{Smo}, \mathit{HRT})$. This means that once we know
age and smoking status of a person and,
for example, that she is not taking $\mathit{HRT}$, then it does not
matter in terms of predicting $\mathit{TCI}$ whether
this is by choice or for instance because $\mathit{HRT}$ is banned from the market.
This
assumption has to be scrutinized with regard to the particular intervention
that is considered and variables that are taken into account.
If, for example, smoking status was unobserved and omitted from
the graph, then the absence of an edge from $\sigma_{\mathit{HRT}}$ to $\mathit{TCI}$ in
the new graph
might not be justifiable as $\mathit{TCI}\perp\!\!\!\perp\hspace
*{-13pt}\diagup\ \sigma_{\mathit{HRT}}|(\mathit{Age}, \mathit{HRT})$ if Figure
\ref{fig_tci2} is correct (see moral graph in Figure \ref{fig_tci2m2}).
We might even doubt the independence $\mathit{TCI}\perp\!\!\!\perp\sigma
_{\mathit{HRT}}|(\mathit{Age},\break \mathit{Smo}, \mathit{HRT})$
in Figure \ref{fig_tci2}, for example, if it is thought that
socioeconomic background
predicts $\mathit{HRT}$ and $\mathit{TCI}$ in a way not captured by $\{\mathit{Age},\mathit{Smo}\}$.

With an influence diagram constructed as above,
the distribution of all variables under an intervention $\sigma_X=x'$
is given by
(\ref{factor}) with the
only modification that $p(x|$pa$(x))$ is replaced by $\delta\{x=x'\}$
due to
(\ref{eq10}). This results in the well-known intervention formula an early
version of which appears in the article by Davis (\citeyear{1984Davis}) (see also
Spirtes, Glymour  and Scheines, \citeyear{1993Spirtes};
Pearl, \citeyear{1993Pearl}).

We want to stress that influence diagrams are \textit{more general} than
causal DAGs which have become a popular tool in epidemiology (Greenland
et al., \citeyear{1999aGreenland}).
The assumptions underlying a causal DAG are equivalent to those represented
in an influence diagram that has intervention nodes
$\sigma_v$ and edges $\sigma_v\rightarrow v$ for \textit{every} node
$v\in V$ in
the
DAG. The absence
of directed edges from $\sigma_v$ to any other variable than $v$
translates for
a causal DAG to the requirement that all common causes of any pair of variables
have to be included in the graph. Hence, readers who are more familiar
with causal
DAGs can think of influence diagrams as causal DAGs (ignoring $\sigma_X$),
but they are then making stronger assumptions.
For a critical view on causal DAGs see the article by Dawid (\citeyear{2010Dawid}).

\subsection{Population Causal Effect}

We give two definitions of causal effects that are relevant for the present
article. They are in the spirit of similar definitions in the
literature (Rubin, \citeyear{1974Rubin};
Robins,  \citeyear{1986Robins}; Pearl, \citeyear{2000Pearl}; Dawid, \citeyear{2002Dawid}).
We formulate them first in terms of distribution and later specify
particular causal parameters.

A population causal effect is some
contrast between the post-intervention distributions
$p(y|\sigma_X=x)$, $x\in\mathcal{X}$, of $Y$ for different interventions,
for example, setting $X$ to $x_1$ as opposed to $x_2$. One could say that
this is a valid target of inference if we contemplate administering
a treatment to the whole population. Most
radically one can say that $X$ has a causal effect on $Y$ if for some
values $x_1\not=x_2\in\mathcal{X}$ the two distributions
$p(y|\sigma_X=x_i)$, $i=1,2$, differ in some aspect. If one
can estimate these post-intervention distributions from observable data,
then one can estimate any contrast between them.
When $p(y|\sigma_X=x)\not=p(y|X=x;\sigma_X=\varnothing)$ we say that
the effect
of $X$ on $Y$ is (marginally)
confounded.\footnote{Note that ``reverse causation'' can occur,
when in fact $Y$ is the cause of $X$, in which case we also have
$p(y|\sigma_X=x)\not=
p(y|X=x;\sigma_X=\varnothing)$. This is relevant in case-control studies,
where it is not always ensured that $X$ is prior to $Y$; for example,
when $Y$ is
coronary heart disease and $X$ is homocysteine level, one might argue that
existing atherosclerosis increases the homocysteine level. We do not consider
reverse causation as confounding.}
[Note that as detailed by Greenland, Pearl and Robins (\citeyear{1999bGreenland}) it is important
to treat confounding and noncollapsibility as distinct concepts.]
We can adjust for confounding if a set of variables $C$ is observed
satisfying the
following conditions (\ref{eq1}) and (\ref{eq4}) [in short we call
this a \textit{sufficient set of covariates}
(Dawid, \citeyear{2002Dawid})].
Assume we know that $Y\perp\!\!\!\perp\sigma_X|(X,C)$, that is,
\begin{eqnarray} \label{eq1}
&&p(y| C=c;\sigma_X=x)\nonumber
\\[-8pt]\\[-8pt]
&&\quad=p(y|X=x, C=c; \sigma_X=\varnothing),\nonumber
\end{eqnarray}
meaning that once we know $C$ and the value $x$, then it does not
make a difference whether $X=x$ has been observed to happen by
nature or by intervention. If in addition
%
\begin{equation}
C\perp\!\!\!\perp\sigma_X, \label{eq4}
\end{equation}
that is, the covariates $C$ are pre-treatment, then the post-intervention
distribution can be consistently estimated from prospective data
(provided $C$ is observed). The post-intervention
distribution for setting $X$ to $x$ is obtained as
\begin{eqnarray} \label{eq2}
&&p(y|\sigma_X=x)\nonumber
\\
&&\quad=\sum_{x',c}p(y|c,x';\sigma_X=x)p(x'|c;\sigma_X=x)p(c)\nonumber
\\[-8pt]\\[-8pt]
&&\quad=\sum_{x',c}p(y|c,x';\sigma_X=\varnothing)\delta\{x=x'\}p(c)\nonumber
\\
&&\quad=\sum_{c}p(y|c,x;\sigma_X=\varnothing)p(c),\nonumber
\end{eqnarray}
where the last step is due to (\ref{eq10}). The quantities
$p(y|c,x;\sigma_X=\varnothing)$ and $p(c)$ can be consistently estimated
from prospective data on $X,Y$ and $C$.
As pointed out, for example, by Clayton (\citeyear{2002Clayton}), (\ref{eq2})
corresponds to
classical direct standardization.
The above conditions (\ref{eq1}) and (\ref{eq4}) are equivalent to
Pearl's (\citeyear{1995Pearl}, \citeyear{2000Pearl}) so-called back-door criterion for causal graphs
(Lauritzen, \citeyear{2000Lauritzen}). If we cannot find a set of covariates that satisfies
(\ref{eq1}) and (\ref{eq4}), an alternative is to use an instrumental
variable,
but we do not consider this any further here (see Angrist, Imbens and Rubin \citeyear{1996Angrist}; Didelez
and Sheehan, \citeyear{2007aDidelez}).

\begin{figure}[t]

\includegraphics{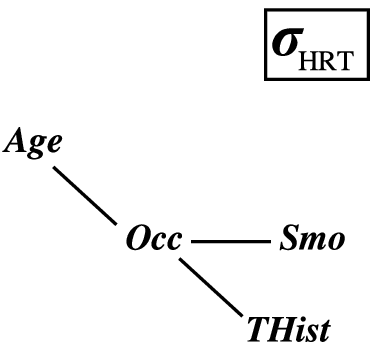}

\caption{Moral graph on $\mathit{Age}$, $\mathit{Occ}$, $\mathit{Smo}$, $\mathit{THist}$
and $\sigma$
for Figure~\protect\ref{fig_tci2}.}
\label{fig_tci2m1}
\end{figure}

\begin{figure}[b]

\includegraphics{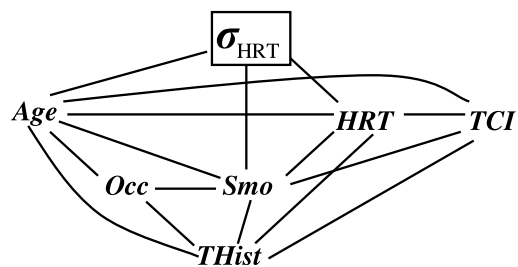}

\caption{Moral graph on all variables
for Figure \protect\ref{fig_tci2}.}
\label{fig_tci2m2}
\end{figure}

\begin{figure}[t]

\includegraphics{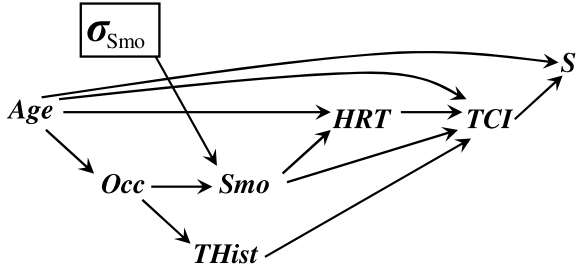}

\caption{Influence diagram for $\mathit{TCI}$ example with intervention
in `smoking.'}
\label{fig_tci_causal2}
\end{figure}

\begin{Example continued*}\hspace*{-3pt}
Consider again
Figure~\ref{fig_tci2}.
We can see that $C=\{\mathit{Age}$, $\mathit{Occ}$, $\mathit{Smo}$, $\mathit{THist}\}$ satisfies (\ref
{eq1}) and (\ref{eq4}).
But these properties are also satisfied for the smaller set $C'=\{\mathit{Age},
\mathit{Smo}\}$.
$\mathit{Age}$ and $\mathit{Smo}$ are independent of $\sigma_{\mathit{HRT}}$, as can be seen from
the moral graph in Figure \ref{fig_tci2m1}, and together with $\mathit{HRT}$
they separate $Y$ and
$\sigma_{\mathit{HRT}}$ as can be seen from the second moral graph in
Figure \ref{fig_tci2m2}.
This implies that in a prospective study we can ignore $\mathit{Occ}$ and
$\mathit{THist}$ altogether and apply
(\ref{eq2}) to obtain the post-intervention distribution.

If, instead, we were to investigate the causal effect of \textit{smoking}
on $\mathit{TCI}$ we might assume an influence diagram as in Figure \ref
{fig_tci_causal2} (ignoring $S$). We can see by a similar reasoning
that $C=\{\mathit{Occ}\}$ is a sufficient set of covariates.
Note that, in this case, the mediating variable $\mathit{HRT}$ must not be
included in
$C$ as it does not satisfy (\ref{eq4}).
This illustrates that the population causal effect that is identified
by conditions
(\ref{eq1}) and (\ref{eq4}) is an overall or total effect, for
example, the effect of
\textit{smoking} on $\mathit{TCI}$ as potentially mediated by its effect on $\mathit{HRT}$.
\end{Example continued*}

As can be seen from (\ref{eq2}), the population causal effect
depends on the distribution of $C$ in the population; this is not
always desirable as it may mean that we cannot carry forward the results
to another population. Hence we consider the conditional causal effect next.

\subsection{Conditional Causal Effects}
A conditional causal effect is some contrast
between the post-intervention distributions conditional on some
covariates $C$,
$p(y|C;\sigma_X=x)$, $x\in\mathcal{X}$ (for the moment $C$
need not be the same as in (\ref{eq2}), but we get back to
this). Such a conditional causal effect may be of interest if one
wants to measure how effective treatment is for a particular patient with
known characteristics such as gender, medical history, etc.
It therefore seems
reasonable to assume that these covariates $C$ satisfy (\ref{eq4}). We
further assume that they also satisfy (\ref{eq1})
because otherwise we would need to take additional suitable
covariates into account in order to apply (\ref{eq2}), so we might
as well incorporate them immediately. Also, if $C$ satisfies both
properties, the conditional causal effect does not depend on
the population distribution of covariates. With (\ref{eq1}) and
(\ref{eq4}) the conditional post-intervention distribution
is automatically identified if $C$ is observed. Note that in order to
obtain the population causal effect using (\ref{eq2}) we can choose any
set $C$ such that (\ref{eq1}) and (\ref{eq4}) are satisfied,
whereas when we consider the conditional causal effect $C$ could include
more variables, for example, because they are so-called effect
modifiers. For example, in Figure \ref{fig_tci2} one may be interested
in the conditional causal effect
given $\mathit{Age}$, $\mathit{Smo}$ {and} $\mathit{THist}$ if the latter is thought
to predict a different effect of $\mathit{HRT}$ on $\mathit{TCI}$, even though it is not
necessary to adjust for $\mathit{THist}$ to obtain the population causal effect.

As alluded to earlier, both the population but also the conditional
causal effect are
``total'' causal effects, when $C$ satisfies (\ref{eq4}), in the sense
that they
include direct as well as indirect effects of $X$ on $Y$; for example,
the effect of
smoking on $\mathit{TCI}$ may be moderated by $\mathit{HRT}$.
A detailed treatment of this topic is beyond the scope of this article
but we refer
to the works of Pearl (\citeyear{2001Pearl}), Robins (\citeyear{2003Robins}) and
Didelez, Dawid and Geneletti (\citeyear{2006Didelez}) and Geneletti (\citeyear{2007Geneletti})
for the general theory, and conditions of identifiability, of direct
and indirect
effects especially in the nonlinear case.

\subsection{Inference on Causal Effect}

We review testing for the causal effect based on prospective data.
In the broadest sense, the causal null hypothesis is that the
post-intervention distribution of $Y$, $p(y|\sigma_X=x)$ (or possibly
$p(y|c;\sigma_X=x)$ if we consider the conditional causal effect),
does not depend on the value $x$,
that is, we do not change the distribution of $Y$ by setting $X$ to
different values.
It is clear from (\ref{eq2}) that if there is no conditional causal
effect, that is, if
$Y\perp\!\!\!\perp X|(C;\sigma_X=\varnothing)$, then there is also no
population causal effect.
The converse is not necessarily true, in particular when there are different
effects in different subgroups that may happen to cancel each other out such
that there is no overall effect in the whole population, that is,
$p(y|\sigma_X=x)$ is independent of $x$ without $Y\perp\!\!\!\perp
X|(C; \sigma_X
=\varnothing)$ being
true---this is known as {lack of} faithfulness (see Spirtes, Glymour and Scheines, \citeyear{1993Spirtes}).
Hence we suggest testing $Y\perp\!\!\!\perp X|(C;\sigma_X=\varnothing
)$ in order to
investigate the causal null
hypothesis of no (conditional) causal effect. If this independence can
be rejected,
then there is evidence for a conditional causal effect, and (except in
rare cases
of such lack of faithfulness) for a population effect.

For estimation, we need to define the causal
parameter of interest. Much of the causal literature is based on the
difference in expectation, leading to the average population
and average conditional causal effect,
$E(Y|\sigma_X=x_1)-E(Y|\sigma_X=x_2)$ (often denoted by $ACE$) and
$E(Y|C=c; \sigma_X=x_1)-E(Y|C=c; \sigma_X=x_2)$, respectively.

Here, we focus instead on population and conditional causal \em odds
ratios \em
as these are invariant to the marginal distributions and hence
applicable under
outcome-dependent sampling, as will be seen. Assume that $Y$ and $X$
are binary.
The \textit{population} causal odds ratio ($\mathit{COR}$) is defined
as
\[
\mathit{COR}_{YX}=\frac{p(Y=1|\sigma_X=1)p(Y=0|\sigma_X=0)}{p(Y=0|\sigma
_X=1)p(Y=1|\sigma_X=0)}.
\]
Alternatively consider the \textit{conditional} $\mathit{COR}_{YX}(C=c)$ where we
condition on the set of covariates $C=c$, that is,
\begin{eqnarray*}
&&\mathit{COR}_{YX}(C=c)\nonumber
\\[-8pt]\\[-8pt]
&&\quad =\frac{p(Y=1|c; \sigma_X = 1) p(Y=0|c; \sigma_X =
0)}{p(Y=0|c; \sigma_X=1)p(Y=1|c;\sigma_X=0)}.\nonumber
\end{eqnarray*}
This is distinct from the population $\mathit{COR}_{YX}$ when it is not
collapsible over $C$,
just as for the associational odds ratio.
When a set of covariates $C$ is sufficient to adjust for confounding,
that is, satisfies (\ref{eq1}) and (\ref{eq4}), then $p(Y=1|c; \sigma
_X=x)=p(Y=1|c, X=x; \sigma_X=\varnothing)$ and hence $\mathit{COR}_{YX}(C)=\mathit{OR}_{YX}(C)$.
This means we can use Corollary \ref{corol_succ}
in order to check whether $C$ can be reduced, that is, whether
$\mathit{OR}_{YX}(C)$ and hence $\mathit{COR}_{YX}(C)$ is collapsible over a subset of $C$.

\subsection{Causal Inference in Case-Control Studies}\label{sec_ccstudies}

\begin{figure}[t]

\includegraphics{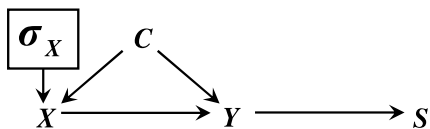}

\caption{DAG representing simple case-control situation.}
\label{fig1}
\end{figure}

Now we include the sampling variable $S$ in our considerations.
Note that the targeted causal parameters do not involve $S$, so we only
want to make assumptions about the distribution of $S$ under the observational
regime $\sigma_X=\varnothing$. In the simple situation of a case-control
study (without matching) sampling is just on the values of $Y$.
Therefore we assume that
%
\begin{equation}\label{s_ind}
S\perp\!\!\!\perp(C,X)|(Y; \sigma_X=\varnothing)
\end{equation}
and in addition
(\ref{eq1}) and (\ref{eq4}). These together imply the following
factorization:
\begin{eqnarray}\label{eq11}
&&p(y,c,x,s|\sigma_X=\varnothing)\nonumber
\\
&&\quad= p(s|y;\sigma_X=\varnothing)p(y|c,x)
\\
&&{}\qquad\times p(x|c;\sigma_X=\varnothing)p(c).\nonumber
\end{eqnarray}
The influence diagram in Figure \ref{fig1} represents slightly stronger
restrictions, as it implies $S\perp\!\!\!\perp\sigma_X|Y$ which does
not follow from
(\ref{eq1}), (\ref{eq4}) and (\ref{s_ind}); that is, we do not
specify any assumptions
about the distribution of $S$ under $\sigma_X=x$ as this is not
relevant to the
target of inference.
We will nevertheless use influence diagrams like Figure \ref{fig1} to represent
jointly our assumptions about the sampling process and the contemplated
intervention.

\begin{figure}[b]

\includegraphics{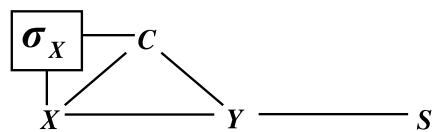}

\caption{Moral graph for simple case-control
situation.} \label{fig2}
\end{figure}

The data come from the distribution $p(y,c,x|S=1;
\sigma_X=\varnothing)$, given by
%
\begin{equation}\label{eq3}
\frac{
p(S=1|y)p(y|c,x)p(x|c;\sigma_X=\varnothing)p(c)}{p(S=1|\sigma
_X=\varnothing)},
\end{equation}
similar to (\ref{eq9}).
The moral graph in Figure \ref{fig2}
includes an edge between $\sigma_X$ and $C$ as the conditional
distribution of $C$ given $S$
is not the same for different regimes $\sigma_X$.
Hence if individuals are selected based on their case or control
status, we cannot expect the distribution of the covariates to be the same
in a scenario where the risk factor $X$ has been manipulated by
external intervention
as in a scenario where it has been left to arise naturally.

The following theorem revisits the well-known result that the causal effect
of $X$ on $Y$ can be tested, and the causal odds ratio estimated,
from case-control data (Breslow, \citeyear{1996Breslow}). The target of
inference is\break $\mathit{COR}_{YX}(C)$, based on $p(y|c;\sigma_X=x)$,
which is \textit{prospective} in the sense that we want to
predict the effect of manipulating $X$ on $Y$ after knowing $C$ without
conditioning on $S$
while we have only the \textit{retrospective} information $p(y|c,S=1;
\sigma_X=\varnothing)$ available.
The following theorem allows $S$ to depend on the covariates $C$
as well as on $Y$.

\begin{theo}\label{theo_prosp}
Under (\ref{eq1}), (\ref{eq4}) and assuming $S\perp\!\!\!\perp
X|(C,Y;\sigma
_X=\varnothing)$, we can
\textup{(i)} test the null hypothesis of no conditional causal effect of $X$ on
$Y$ given $C$ by testing $X\perp\!\!\!\perp Y|(C,S=1;\sigma
_X=\varnothing)$
(regardless of the measurement scales), and
\textup{(ii)} consistently estimate $\mathit{COR}_{YX}(C)$ by estimating
$\mathit{OR}_{YX}(C,S=1)$ (for categorical $X,Y$).
\end{theo}

\begin{pf}
(i)
Earlier we argued that a test for\break $Y\perp\!\!\!\perp
X|(C;\sigma=\varnothing)$ can replace a test of the null hypothesis of
no causal
effect when $C$ satisfies (\ref{eq1}) and (\ref{eq4}).
As $S\perp\!\!\!\perp X|(C,Y;\sigma_X=\varnothing)$, Theorem \ref
{null_test} completes
the argument.

(ii)
Assumptions (\ref{eq1}) and (\ref{eq4}) imply $\mathit{COR}_{YX}(C)=\mathit{OR}_{YX}(C)$
as explained earlier. With Corollary \ref{corol_retro} we see that
$\mathit{OR}_{YX}(C,S)$ is collapsible over $S$ when
$S\perp\!\!\!\perp\break
X|(C,Y;\sigma
_X=\varnothing)$,
which completes the proof.
\end{pf}

In Theorem \ref{theo_prosp}, as far as testing is concerned,
we are not restricted to the categorical situation and can use as test
statistic whatever seems appropriate given the measurement scales of
$X,Y,C$. If this independence is rejected, then there is evidence for a
causal effect. In the particular case of binary $Y$ and continuous $X$
it is well known that we can still also consistently estimate the odds
ratio using a logistic regression (Prentice and Pyke, \citeyear{1979Prentice}).
Their result, however, relies on the logistic link being justified,
while the results on odds ratios when $X$ and $Y$ are both categorical,
such as Theorem~\ref{theo_prosp}(ii),
make no parametric assumptions.

The set $C$ in Theorem \ref{theo_prosp} needs to contain a sufficient set
of covariates so as to justify (\ref{eq1}). But it also needs to
contain any
matching variables, even if these are not needed for (\ref{eq1}),
in order to justify $S\perp\!\!\!\perp X|(C,Y;\sigma_X=\varnothing)$.
This has been
illustrated
in Figure \ref{matched}, with the variable $B$ which is not needed to
adjust for
confounding. Hence, a sufficient set
of covariates and the matching variables are required
for Theorem \ref{theo_prosp} to work. However, typically a much larger
set of
covariates has been observed;
one can then use Corollary \ref{cor1} to reduce it
without losing information, as in the following example.

\begin{Example continued*}
In the $\mathit{HRT}$--$\mathit{TCI}$ example, as the study design was case-control
matched by
$\mathit{Age}$, we need to make sure that $C$ contains $\mathit{Age}$. But we already saw that
$C=\{\mathit{Age}, \mathit{Smo}\}$ is a set of sufficient covariates. Hence, all assumptions
of Theorem \ref{theo_prosp} are satisfied with this choice of $C$
(check these on the influence diagram in Figure
\ref{fig_tci_causal}). That is, we can consistently estimate the
causal odds ratio
between $\mathit{HRT}$ and $\mathit{TCI}$ given $\mathit{Age}, \mathit{Smo}$ from the available data.

Alternatively, if the target is the conditional causal odds ratio given
\textit{all}
covariates, then we can see that with the choice of $C'=\{\mathit{Age}, \mathit{Occ},
\mathit{THist}, \mathit{Smo}\}$
the conditions of Theorem \ref{theo_prosp} are satisfied; we can
estimate the
causal odds ratio $\mathit{HRT}$ and $\mathit{TCI}$ given $C'$ from the available data,
but we can
additionally omit $\mathit{Occ}$ due to the conditional
odds ratio being collapsible over this variable.
Note that it is not further collapsible over the variable $\mathit{THist}$, implying
that the causal odds ratio given $C=\{\mathit{Age}, \mathit{Smo}\}$ is different from the causal
odds ratio given $C'=\{\mathit{Age}, \mathit{Occ}, \mathit{THist}, \mathit{Smo}\}$, though both
conditioning sets are sufficient
to adjust for confounding under our assumptions.
As mentioned before, $\mathit{THist}$ could be an effect modifier and
might therefore be included.

\begin{figure}[t]

\includegraphics{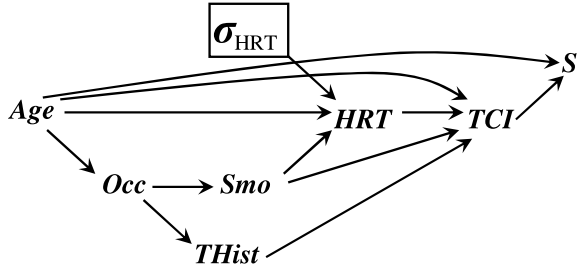}

\caption{Influence diagram for $\mathit{TCI}$ example with matched
sampling.}
\label{fig_tci_causal}
\end{figure}

\begin{table}[t]
\caption{Conditional causal odds ratio between $\mathit{Smo}$ and $\mathit{TCI}$
given ($\mathit{Age}$, $\mathit{THist}$)}
\label{log-or-table4}
\begin{tabular*}{\tablewidth}{@{\extracolsep{4in minus 4in}}lcc@{}}
\hline
\textbf{Smoking level} & \textbf{log-\textit{OR} (stdev)} & \textbf{\textit{OR}}\\
 \hline
Never & Reference & \\
Former & 0.41 (0.21) & 1.51\\
1--10 & 0.89 (0.19) & 2.43\\
11--20 & 0.97 (0.18) & 2.65\\
21$+$ & 1.19 (0.41) & 3.29\\
\hline
\end{tabular*}
\end{table}

Assume now that we are instead interested in the effect of smoking
($\mathit{Smo}$) on $\mathit{TCI}$ and that the assumptions encoded in Figure \ref
{fig_tci_causal2}
are satisfied. So far we have targeted the conditional causal odds
ratio between
exposure and response given \textit{all} covariates; however, if we include
the mediator $\mathit{HRT}$ into $C$, then it does not satisfy condition (\ref
{eq4}) as it
is a descendant of $\mathit{Smo}$. Hence we could
consider $C=\{\mathit{Age}$, $\mathit{Occ}$, $\mathit{THist}\}$ and find that $\mathit{OR}_{\mathit{Smo},\mathit{TCI}}(C,S)$
can again be
collapsed over $\mathit{Occ}$.
The resulting estimates are shown in Table \ref{log-or-table4}.
Note that these describe the ``total'' effect of $\mathit{Smo}$ on $\mathit{TCI}$
including possible mediation via $\mathit{HRT}$ (but conditional on $\mathit{Age}$ and $\mathit{THist}$).
\end{Example continued*}

\section{Extensions and further examples}\label{sec_ext}

In this section we consider more general data situations where the
sampling depends
in a less obvious way on the outcome and possibly on further variables.
In particular, we extend the previous results to the case where a
sufficient set of
covariates (and possibly matching variables) $C$ does not allow us to
collapse over
$S$. In such cases taking further variables into account can sometimes
provide a
solution. Let us start with an example.

\begin{Example*}
Weinberg, Baird and Rowland (\citeyear{1993Weinberg}) and Slama et al. (\citeyear{2006Slama}) considered
`time-to-pregnancy' studies which are of interest when investigating
factors affecting fertility.
Typically $X$ is exposure, such as a toxic substance or smoking, and
$Y$ is the time to pregnancy; common covariates $C$ such as age,
socioeconomic background, etc., may be taken into account.
The problem here is that if women are sampled who became pregnant
during a certain time interval (retrospective sampling), then
long duration to pregnancy automatically means earlier initiation time.
However, initiation time might predict the exposure if it has changed
over time, for example, because precautions regarding toxic substances
have increased or smoking habits in the population have changed over time.
Therefore, $Y$ and $X$ may be associated given $C$ even if there is no
causal effect of exposure, that is, $C$ is not ``bias-breaking.''
Note that the same phenomenon also occurs with current duration designs
and that prospective
sampling has many other drawbacks in ``time-to-pregnancy'' studies as
discussed in
detail by Slama et al. (\citeyear{2006Slama}).

The key to solving this problem is to find a bias-breaking variable $Z$
such that either $X$ or $Y$ can reasonably be assumed conditionally
independent of
$S$ given $Z$ (and observed covariates) and to use Corollary \ref
{corol_succ} so
as to further collapse over $Z$. The method proposed by Weinberg, Baird and
Rowland (\citeyear{1993Weinberg}) relies on using the time of initiation. It seems plausible
that once the initiation time $Z$ and time to pregnancy $Y$ are known,
the sampling $S$ is not further associated with the exposure $X$,
typically controlling for relevant covariates~$C$, that is, $S\perp\!\!
\!\perp X|(Z,Y,C)$.
Further, we may sometimes be able to justify that $Y\perp\!\!\!\perp
Z|(X,C)$, that
is, that the initiation time itself, once we account for relevant
factors and regardless of whether the unit is sampled or not, should
not predict time to pregnancy.
This assumption might be violated if there are other relevant factors
that have changed over time and that are not captured by $C$ or $X$.
\end{Example*}

\begin{figure}[t]

\includegraphics{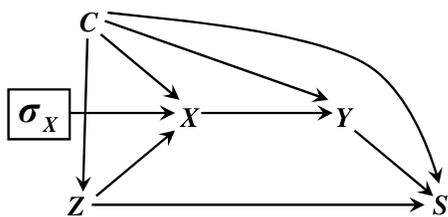}

\caption{Graphical representation of assumption in time-to-pregnancy example.}
\label{fig4}
\end{figure}

Using again $\sigma_X$ as intervention indicator and assuming that $C$
is a sufficient set of covariates, we can summarize our
assumptions about the time-to-pregnancy example through the following
independencies: $Z\perp\!\!\!\perp\sigma_X$, $Y\perp\!\!\!\perp
(Z,\sigma_X)|(X, C)$ and $S\perp\!\!\!\perp\break (X,\sigma_X
)|(Z,Y,C)$.
The graph in Figure \ref{fig4} represents these conditional independence
assumptions (the edge $C\rightarrow Z$ could be replaced by
$C\leftarrow Z$).
If it were not for the retrospective
sampling, the causal effect of $X$ on $Y$ could be analyzed ignoring
$Z$, as $C$ is assumed a sufficient set of covariates.
The selection effect becomes apparent when checking for graph
separation, which yields a moral edge between $Z$ and $Y$ when
conditioning on $S$
(cf. moral graph in Figure \ref{fig4b2}).

The following theorem shows that exploiting the initiation time $Z$ in
the above example can indeed facilitate inference about the causal effect.

\begin{theo}\label{theo1}
We can test for a (conditional) causal effect of $X$ on $Y$ (given $C$)
if there
exists a set of observable variables $Z$ such that all following
conditions are satisfied:
\begin{longlist}[(iii)]
\item[(i)] $S\perp\!\!\!\perp X|(Y,Z,C;\sigma_X=\varnothing)$,
\item[(ii)] $Y\perp\!\!\!\perp Z|(X,C;\sigma_X=\varnothing)$,
\item[(iii)] $C$ is a sufficient set of covariates,
\item[(iv)] the joint distribution $p(y,x,z|\sigma_X=\varnothing)$ is
stri\-ctly positive.
\end{longlist}
The causal null hypothesis is then equivalent to $Y\perp\!\!\!\perp
X|(Z,C,S=1;\sigma_X=\varnothing)$.
\end{theo}

\begin{pf}
As argued before, testing $Y\perp\!\!\!\perp X|(C;\break \sigma_X=\varnothing
)$ provides a test
for the causal
null hypothesis. We show that it is equivalent to $Y\perp\!\!\!\perp
X|(Z,C, S=1;
\sigma_X=\varnothing)$. As all conditional independencies are
conditional on
$\sigma_X=\varnothing$
it will be omitted from the notation.

Remember that with (i) and Theorem \ref{null_test}, $Y\perp\!\!\!
\perp X|(Z,C)$ is
equivalent to
$Y\perp\!\!\!\perp X|(Z,C,S=1)$.
First assume that $Y\perp\!\!\!\perp X|C$. With (ii) we obtain $Y\perp
\!\!\!\perp X|(Z,C)$,
which is
equivalent to $Y\perp\!\!\!\perp X|(Z,C,S=1)$. For the converse,
assume $Y\perp\!\!\!\perp
X|(Z,C,S=1)$, hence
$Y\perp\!\!\!\perp X|(Z,C)$. Now, (iv) is sufficient to ensure
(Lauritzen, \citeyear{1996Lauritzen},
page 29)
that this conditional independence together with (ii) yields $Y\perp\!
\!\!\perp(X,Z)|C$
which implies $Y\perp\!\!\!\perp X|C$.
\end{pf}

\begin{figure}[t]

\includegraphics{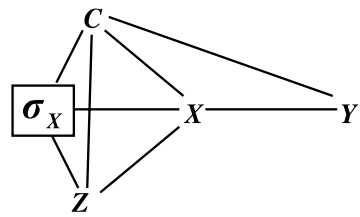}

\caption{Moral graph for DAG in Figure \protect\ref{fig4} marginal over~$S$.}
\label{fig4b1}
\end{figure}

\begin{figure}[b]

\includegraphics{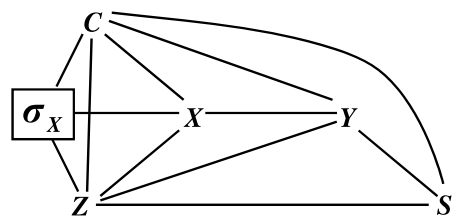}

\caption{Moral graphs for the DAG in Figure \protect\ref{fig4} including~$S$.}
\label{fig4b2}
\end{figure}

\begin{Example continued*} Consider again the time-to-pregnancy study
represented in
Figure \ref{fig4}.
We see that all assumptions of Theorem \ref{theo1} are satisfied:
$C$ and $\sigma$ are nondescendants of each
other and have no parents; from the moral graph on
$Y,Z,X,C$ (Figure \ref{fig4b1}) it follows that $C$ is a sufficient set
of covariates, and that (ii) holds, while part (i) can be seen from the
moral graph on all nodes in Figure \ref{fig4b2}.
Hence we can investigate the null hypothesis of no causal effect by testing
whether $Y$ and $X$ are \em associated \em conditionally on $Z$ and $C$
for the sampled subjects.
\end{Example continued*}

\begin{remarks*} (a)
Theorem \ref{theo1} is symmetric in $X,Y$, in the sense that they
can be swapped in (i) and (ii).

(b) If $Z=\varnothing$, the conditions are the same as
for the matched case-control situation of Section \ref{sec_ccstudies} (Theorem
\ref{theo_prosp}).

(c) Further, the theorem does not require that\break
$Z\perp\!\!\!\perp \sigma_X$, that is, that the bias-breaking
variable $Z$ is not
affected by an
intervention in $X$, hence in Figure \ref{fig4} the arrow from $Z$ to
$X$ could be
reversed (though this is not plausible in the time-to-pregnancy
scenario that we
have used, but may be relevant in other scenarios).
\end{remarks*}

It is easy to find realistic examples where the assumptions of Theorem
\ref{theo1}
are not satisfied; for example, Robins (\citeyear{2001Robins}) explained why it is
difficult to identify the effect of hormone treatment on endometrial
cancer in case-control
studies. In such situations, additional outside information can
sometimes be used
to obtain identifiability (see, e.g., Geneletti, Richardson and Best, \citeyear{2009Geneletti}).

In addition to the above result about testing the causal effect, we
also have the following about estimating it when all variables are discrete.

\begin{theo}\label{theo5}
Under the assumptions of Theorem~\ref{theo1},
we can consistently estimate the (conditional) causal odds ratio of $X$
on $Y$ given $C$, $\mathit{COR}_{YX}(C)$,
by estimating $\mathit{OR}_{YX}(Z,C,S=1)$.
\end{theo}

\begin{pf}
From (iii) it follows that $\mathit{COR}_{YX}(C)=\mathit{OR}_{YX}(C)$, that is, the causal
odds ratio is equal to the observational odds ratio.
Further, using Corollary~\ref{corol_succ} with $B_1=S$ and $B_2=Z$,
(i) yields
$\mathit{OR}_{YX}(Z,\break C,S)$ is collapsible over $S$ in the observational regime
and (ii) means $\mathit{OR}_{YX}(Z,C)$ is collapsible over $Z$, hence $\mathit{OR}_{YX}(Z,C,S)$
is collapsible over $(Z,S)$.
\end{pf}

\begin{remarks*}
(a) In the situation of Theorem \ref{theo5}
it does not necessarily hold that $\mathit{OR}_{YX}(C,S=1)=\mathit{OR}_{YX}(C)$,
that is, Corollary \ref{cor1} (with $B=Z$) does not apply as neither
part (i) nor part (ii)
of that corollary is satisfied.
However, Corollary \ref{cor1} can be used to further reduce
the set $C$ if $\mathit{OR}_{YX}(C)$ is collapsible over a subset of $C$.
The difference between the above Theorem \ref{theo5} and Corollary
\ref{cor1} is
that while
the latter focuses on such a reduction of dimensionality,
the former exploits the fact that
$\mathit{OR}_{YX}(C)$ is the same in a higher dimensional model
$\mathit{OR}_{YX}(C,\break Z,S=1)$.
This is useful because we can only\break estimate quantities conditional on $S=1$
while\break $\mathit{OR}_{YX}(C)\not=\mathit{OR}_{YX}(C,S=1)$.

(b) Theorem \ref{theo5} implies that the assumptions of Theorem
\ref{theo1} can be tested to a certain
extent as they imply that $\mathit{OR}_{YX}(Z=z,C,S=1)=\mathit{OR}_{YX}(Z=z',\break C,S=1)$ for
$z\not= z'$.
Hence, estimates should not vary much for different values of
$Z$.
\end{remarks*}

We conclude with an example for potential selection bias that is not
due to
outcome-dependent sampling but is also covered by the conditions of
Theorem \ref{theo5} due to their symmetry in $X$ and $Y$.

\begin{figure}[t]
\begin{tabular}{cc}

\includegraphics{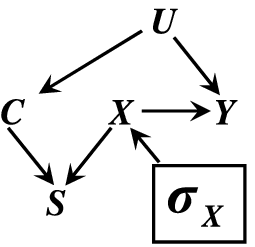}
&\includegraphics{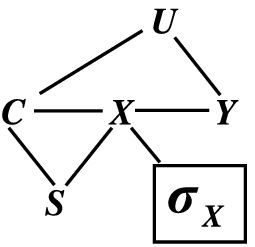}\\
\scriptsize{(a)}&\scriptsize{(b)}\\[6pt]

\includegraphics{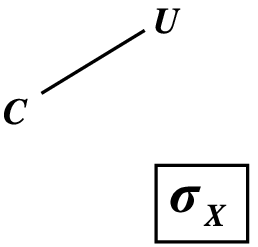}
&\includegraphics{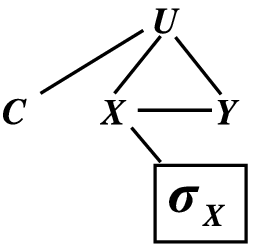}\\
\scriptsize{(c)}&\scriptsize{(d)}
\end{tabular}
\caption{\textup{(a)} DAG and relevant moral graphs for Hernan, Hern\'andez-D\'iaz
and Robins (\protect\citeyear{2004Hern})
example; \textup{(b)} is moral graph on all variables, \textup{(c)} on $C$ and $\sigma
_X$, \textup{(d)}
on $Y,X,C$ and $\sigma_X$.} \label{fig_hernan}
\end{figure}

\begin{Example*} Hernan, Hern\'andez-D\'iaz and Robins (\citeyear{2004Hern}) considered the example
illustrated in Figure \ref{fig_hernan}. In a study with HIV patients
$X$ is anti-retroviral therapy, $Y$ is AIDS, $U$ is the
true level of immunosupression and $C$ is a collection of symptoms as
well as
measurements on CD4 counts. Further covariates would typically be
included but
for simplicity we omit them here.
We assume $X$ is randomized so it has no graph parents.
The fact that $S$ depends on $C$ and $X$ represents
that patients with worse symptoms and side effects, predicted by
treatment and
baseline covariates, are more likely to drop
out and not be available for the analysis. (Note that if it was
not for having to condition on $S=1$, then we could estimate the population
causal effect of $X$ on $Y$ without further adjustment.)
We can verify that the odds ratio between $X$ and $Y$ is not
collapsible over $S$ from Figure \ref{fig_hernan}(b): neither $X\perp
\!\!\!\perp
S$ nor $Y\perp\!\!\!\perp S$.
However, if we consider the conditional causal effect of
$X$ on $Y$ given $C$, then we can collapse over $S$.
With $Z=\varnothing$, all conditions of Theorem \ref{theo5} with $X$
and $Y$
interchanged are satisfied, so we can estimate $\mathit{COR}_{YX}(C)$
by using $\mathit{COR}_{YX}(C,S=1)$; condition (i) can be seen from the moral
graph in
Figure \ref{fig_hernan}(b), condition (ii) is redundant, condition (iii)
can be seen from Figure \ref{fig_hernan}(c) and (d).
\end{Example*}

\section{Conclusion}\label{sec_disc}

As the sampling or selection mechanism can often create complications
and bias in
statistical analyses, we argued in Section \ref{subsec_graphs_ods}
that the basic assumptions about the sampling,
in terms of conditional independence, should be made
explicit using graphical models including a node for the binary
sampling indicator.
We demonstrated how this allows us to characterize, with simple
graphical rules,
situations in which we can collapse the (conditional) odds ratio over
$S$ (Corollary~\ref{corol_retro}) or,
more generally, when we can test for a (conditional) association
(Theorem \ref{null_test}).
Addressing specifically causal inference, Theorem \ref{theo_prosp}
specifies the \mbox{additional} assumptions required to test for a causal
effect or estimate a (conditional) causal odds ratio under
outcome-dependent sampling, such as in a matched case-control design.
Theorems \ref{theo1} and \ref{theo5} extend these results to more
general situations with less obvious outcome-dependent sampling.
Our results are therefore relevant to a range of study designs,
case-control being the most common,
but also, for example, retrospective sampling that is conditional on
reaching a
certain state, such as time-to-pregnancy studies.

We have shown how different types of graphical models can be used to express
assumptions about the sampling process, admitting more flexibility than if
restricted to causal DAGs (but as explained in Section~\ref{ID_CDAG}
our results
are also valid for the latter). In addition to directed acyclic and
undirected graphs, we want
to point out that chain graphs provide a further class of useful
models. The
original analysis of the $\mathit{TCI}$ data, for instance, used chain graphs
(Pedersen et al., \citeyear{1997Pedersen}). The causal interpretation of chain graphs,
however, is more complicated than
for DAGs (cf. Lauritzen and Richardson, \citeyear{2002Lauritzen}).

As any type of graph only encodes presence or absence of
conditional independencies, it cannot represent particular parametric
assumptions or properties of the model and selection process.
Consequently, any inference other than testing or estimating odds ratios
will typically require such additional assumptions, which in turn will
need to be
scrutinized and complemented by a sensitivity analysis.
We therefore regard the use of graphical models in this context
as an important first step of the analysis,
facilitating the structuring and reasoning about the problem of
outcome-dependent
sampling.

Concerning the question of causal inference,
we have mainly assumed an approach of
adjusting for confounding by conditioning on suitable covariates in the
analysis.
A different way of using covariates is via the propensity score
(Rosenbaum and
Rubin, \citeyear{1983Rosenbaum}) or inverse probability weighting (Robins, Hernan and Brumback, \citeyear{2000Robins}),
but little is known as yet on how to adapt these to case-control
studies or
general outcome-dependent sampling; but see the work of Robins, Rotnitzky and Zhao
(\citeyear{1994Robins}, Section 6.3), Newman (\citeyear{2006Newman}),
Mansson et al. (\citeyear{2007Mansson}) and van der Laan (\citeyear{2008van}).

\section*{Acknowledgments}

This work was initiated while Vanessa Didelez and Niels Keiding were
working on ``Statistical Analysis of Complex Event History Data'' at
the Centre for Advanced Study of the Norwegian Academy of Science and
Letters in Oslo, 2005/2006.

\end{document}